\documentclass[ALICE,manyauthors]{cernphprep}

\usepackage[comma,square,numbers,sort&compress]{natbib}
\usepackage{hyperref}
\usepackage{lineno}
\modulolinenumbers[5]
\usepackage{color}
\usepackage[T1]{fontenc}

\newcommand{\snn}{\ensuremath{\sqrt{s_{\mathrm{\scriptscriptstyle NN}}}}\xspace}

\newcommand{\Rz}{\ensuremath{\rho^{0}}\xspace}

\begin{document}%

\begin{titlepage}
\PHyear{2020}
\PHnumber{021}      
\PHdate{24 February}  

%

\title{Coherent photoproduction of $\mathbf{\rho^{0}}\xspace$ vector mesons in ultra-peripheral Pb--Pb collisions at  $\mathbf{\sqrt{\textit{s}_{\rm {\scriptscriptstyle \mathbf{NN}}}} = 5.02}$~TeV}

\ShortTitle{$\Rz$ photoproduction in  Pb--Pb UPC at  $\snn = 5.02$~TeV}   

\Collaboration{ALICE Collaboration\thanks{See Appendix~\ref{app:collab} for the list of collaboration members}}
\ShortAuthor{ALICE Collaboration} 

\begin{abstract}
Cross sections for the coherent photoproduction of $\rho^{0}$ vector mesons  in ultra-peripheral Pb--Pb collisions at $ \sqrt{s_{\mathrm{NN}}}= 5.02$~TeV are reported. The measurements, which rely on the $\pi^+\pi^-$ decay channel,  are presented in three regions of rapidity covering  the range $|y|<0.8$. For each rapidity interval, cross sections are shown for different nuclear-breakup classes defined according to the presence of  neutrons measured in the zero-degree calorimeters. The results are compared  with predictions based on different models of nuclear shadowing. Finally, the observation of a coherently produced resonance-like structure with a mass around 1.7~GeV/$c^2$ and a width of about 140~MeV/$c^2$ is reported and compared with similar observations from other experiments.
\end{abstract}

\end{titlepage}
\setcounter{page}{2}

\section{Introduction
\label{sec:Intro}}
The electromagnetic field of a fast charged particle, such as those circulating in the Large Hadron Collider (LHC), is strongly Lorentz-contracted and its strength is dominated by the component perpendicular to the direction of motion, such that it can be described as a flux of quasi-real photons. The intensity of this photon flux is proportional to the square of the electric charge of the particle; thus when lead ions circulate in the LHC there are, in addition to the standard hadronic collisions, also copious photonuclear interactions. 
Ultra-peripheral collisions (UPC) are defined as those for which the impact parameter  is larger than the sum of the radii of the incoming particles, in which case the occurrence of hadronic processes is strongly suppressed due to the short range nature of quantum chromodynamics (QCD), and photon-induced processes dominate the interaction rate. The physics of UPC and recent results obtained at the LHC are reviewed in~\cite{Baltz:2007kq,Contreras:2015dqa}. 

The  photonuclear production of a $\Rz$ vector meson in Pb--Pb UPC at the LHC is particularly interesting, because  its large cross section makes it a good tool to study the approach to the black-disk limit of QCD~\cite{Frankfurt:2002wc}. This process can be pictured as follows: a quasi-real photon, emitted by one of the Pb ions, fluctuates into a QCD object which then interacts elastically either with the other lead nucleus (coherent interaction) or with one of its nucleons (incoherent interaction) and produces a $\Rz$ vector meson. The QCD object can be taken as a vector meson~\cite{Bauer:1977iq}, as a quark-antiquark colour dipole~\cite{Nikolaev:1990ja,Nikolaev:1991et,Mueller:1989st}, or one could consider intermediate diffractive hadronic states as done in the Gribov-Glauber approach~\cite{Frankfurt:2015cwa}.
In these processes, the mean transverse momentum of the produced vector meson is related to the size of the target in the impact parameter plane by a Fourier transformation; hence, it is  restricted to be in the order of 60 (500) MeV/$c$  for coherent (incoherent) interactions.  In the coherent case the target nucleus remains intact, but in UPC of heavy nuclei the photon fluxes are so intense that further photon exchanges between the same nuclei may occur  independently of the production of the vector meson and produce neutrons at beam rapidities due to electromagnetic excitation of one or both of the incoming nuclei~\cite{Baltz:2002pp}. The experimental signature of coherent $\Rz$ photonuclear production is then the presence  of a single $\Rz$ vector meson with fairly low transverse momentum in the detector, accompanied sometimes by one or few neutrons at beam rapidities. 

The coherent photonuclear production of  a $\Rz$ vector meson at midrapidity was extensively studied in Au--Au UPC  at the Relativistic Heavy Ion Collider (RHIC) at three different centre-of-mass energies per nucleon pair $\snn = 62.4$~GeV~\cite{Agakishiev:2011me}, $\snn =130$~GeV~\cite{Adler:2002sc}, and $\snn =200$~GeV~\cite{Abelev:2007nb,Adamczyk:2017vfu}. It was also  studied by ALICE at the LHC in Pb--Pb UPC at $\snn = 2.76$~TeV~\cite{Adam:2015gsa}. 

A model based on a Glauber description~\cite{Frankfurt:2002wc} predicts cross sections  twice larger than those measured at energies of 200~GeV~\cite{Abelev:2007nb}   and  2.76~TeV~\cite{Adam:2015gsa} even though it is compatible with lower-energy  data~\cite{Agakishiev:2011me,Adler:2002sc}. The STARlight model~\cite{Klein:1999qj,Klein:2016yzr}, which is also based on a Glauber-like eikonal formalism, but does not take into account the elastic part of the elementary $\Rz$--nucleon cross section, successfully describes all the data mentioned above. The inclusion of photon inelastic diffraction into large-mass intermediate hadronic states within the Gribov-Glauber framework of nuclear shadowing provides a better comparison with data than the model based only on a Glauber description~\cite{Frankfurt:2015cwa}.
 Nonetheless, the photoproduction of  $\Rz$ off nuclei  is not yet satisfactorily described in all of its aspects and new measurements, particularly at higher energies, are needed to gain a better understanding.

This article reports the first measurement of coherent photonuclear production of $\Rz$ vector mesons  in Pb--Pb UPC at $\snn = 5.02$ TeV. The measurement was performed by the ALICE Collaboration with data recorded in the 2015 Pb--Pb run. The cross section for this process is measured as a function of the rapidity of the vector meson ($y$) in the range $|y|<0.8$. At each rapidity, the cross sections are reported for the following nuclear-breakup  classes defined by the appearance of neutrons at beam rapidities: 0n0n (no  neutrons), 0nXn (neutrons are measured only on one beam side, either at positive or negative rapidity), and XnXn (neutrons are detected in both beam directions). In the following, they are denoted in general as forward-neutron classes. 
 Furthermore, the observation of a resonance-like structure  in the $\pi^+\pi^-$ invariant mass spectrum at a mass around 1.7~GeV/$c^2$ is reported and compared with similar observations from other experiments.

\section{Experimental set-up
\label{sec:Setup}}
The analysed data were recorded by ALICE towards the end of 2015 when the LHC provided 
Pb--Pb collisions at  $\snn = 5.02$~TeV. A full description of  ALICE systems is given in~\cite{Aamodt:2008zz} and the performance of the detector is discussed in~\cite{Abelev:2014ffa}. Here, only the components relevant for the analysis are briefly described.
The $\Rz$ meson is reconstructed  through its decay into a $\pi^+\pi^-$ pair using the Inner Tracking System (ITS) and the Time Projection Chamber (TPC) to measure the pion tracks. Vetoes on the presence of other particles to ensure that  only the $\Rz$ meson is produced are imposed with the V0 and the ALICE Diffractive (AD) detectors. The neutrons at beam rapidities are measured with the Zero Degree Calorimeters (ZDC).

The ITS~\cite{Aamodt:2010aa} is the innermost detector system of ALICE. It consists of six cylindrical layers of silicon detectors, positioned coaxially with the direction of the incoming beams, which defines the $z$-axis. This detector covers the full azimuthal angle and the pseudorapidity range $|\eta|<0.9$. All six layers contribute to track reconstruction. 
 The Silicon Pixel Detector (SPD) makes up the first two layers of the ITS, closest to the beam, and is
  particularly important for this analysis because it participates in the trigger definition. The SPD has $9.8\times10^6$ pixels of reverse-biased silicon diodes, which are read out by 400 (800) chips in the inner (outer) layer. 
Each of the readout chips fires a trigger  if at least one of its pixels has a signal. When projected into the transverse plane, the chips define 20 (40) azimuthal regions in the inner (outer) layer.

The TPC~\cite{Alme:2010ke} is the main tracking detector. It is a large cylindrical gas detector with a  central membrane at high voltage and readout planes, composed of multi-wire proportional chambers, at each of the two end caps. It covers the full azimuthal range and $|\eta|<0.9$ for tracks which fully traverse it. It provides up to 159 space points for track reconstruction and for particle identification by measuring the ionisation energy loss. Both the ITS and the TPC are inside a large solenoid magnet, which creates  a  uniform 0.5~T magnetic field parallel to the $z$-axis.

The V0~\cite{Abbas:2013taa}  is a set of two segmented scintillator counters, V0A and V0C.
The V0A covers the range $2.8<\eta<5.1$, while the V0C covers $-3.7 < \eta < -1.7$.  The AD~\cite{N.Cartiglia:2015gve} is also a set of two arrays of scintillator detectors, ADA and ADC, placed further away from the nominal interaction point and covering $4.7<\eta<6.3$ and $-6.9<\eta<-4.9$, respectively. 
Both V0 and AD detectors participate in the first level trigger, and both detectors have timing resolution less than 1 ns.

\begin{figure}[t!]
\centering
 \includegraphics[width=0.48\textwidth]{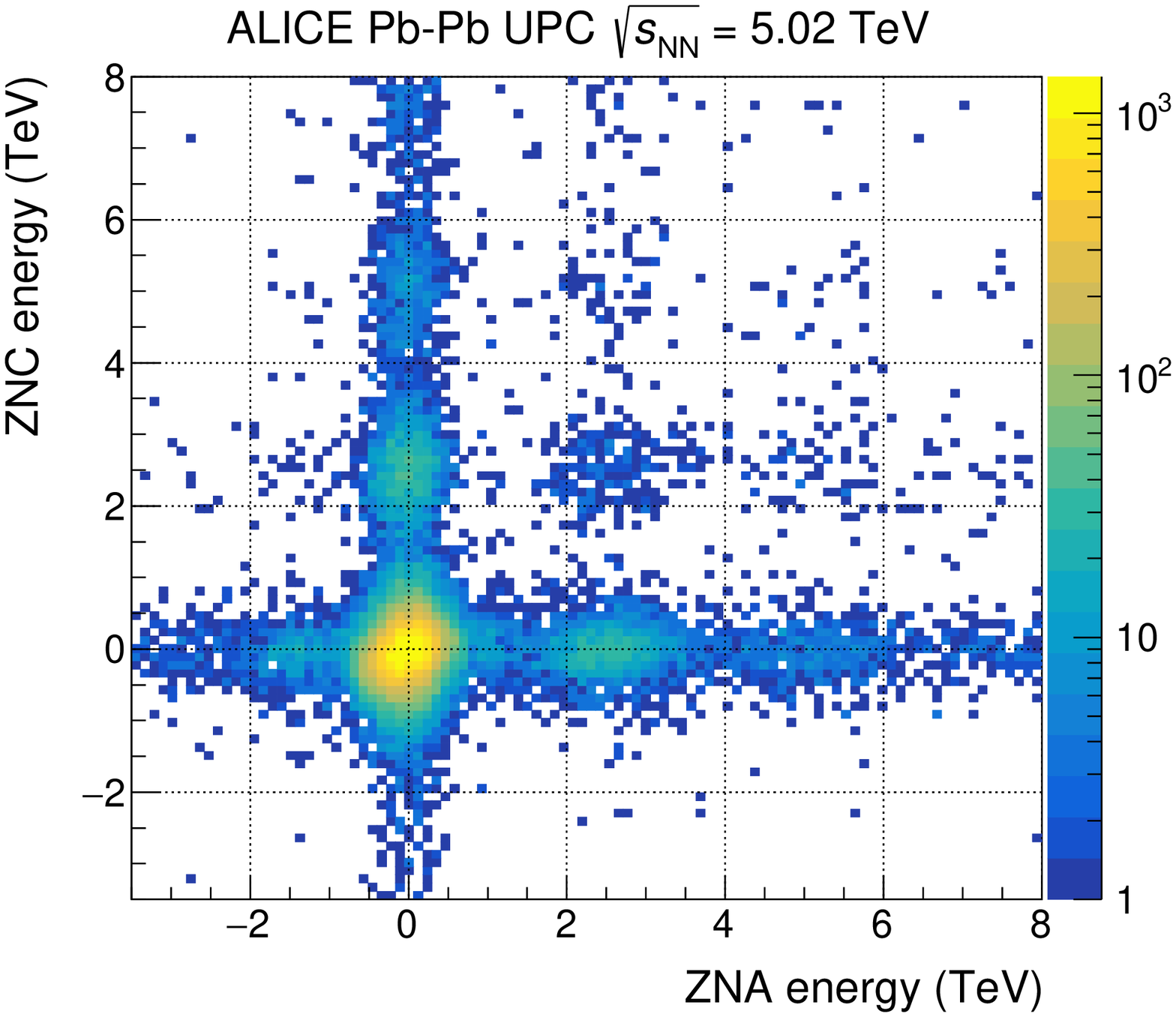}
  \includegraphics[width=0.48\textwidth]{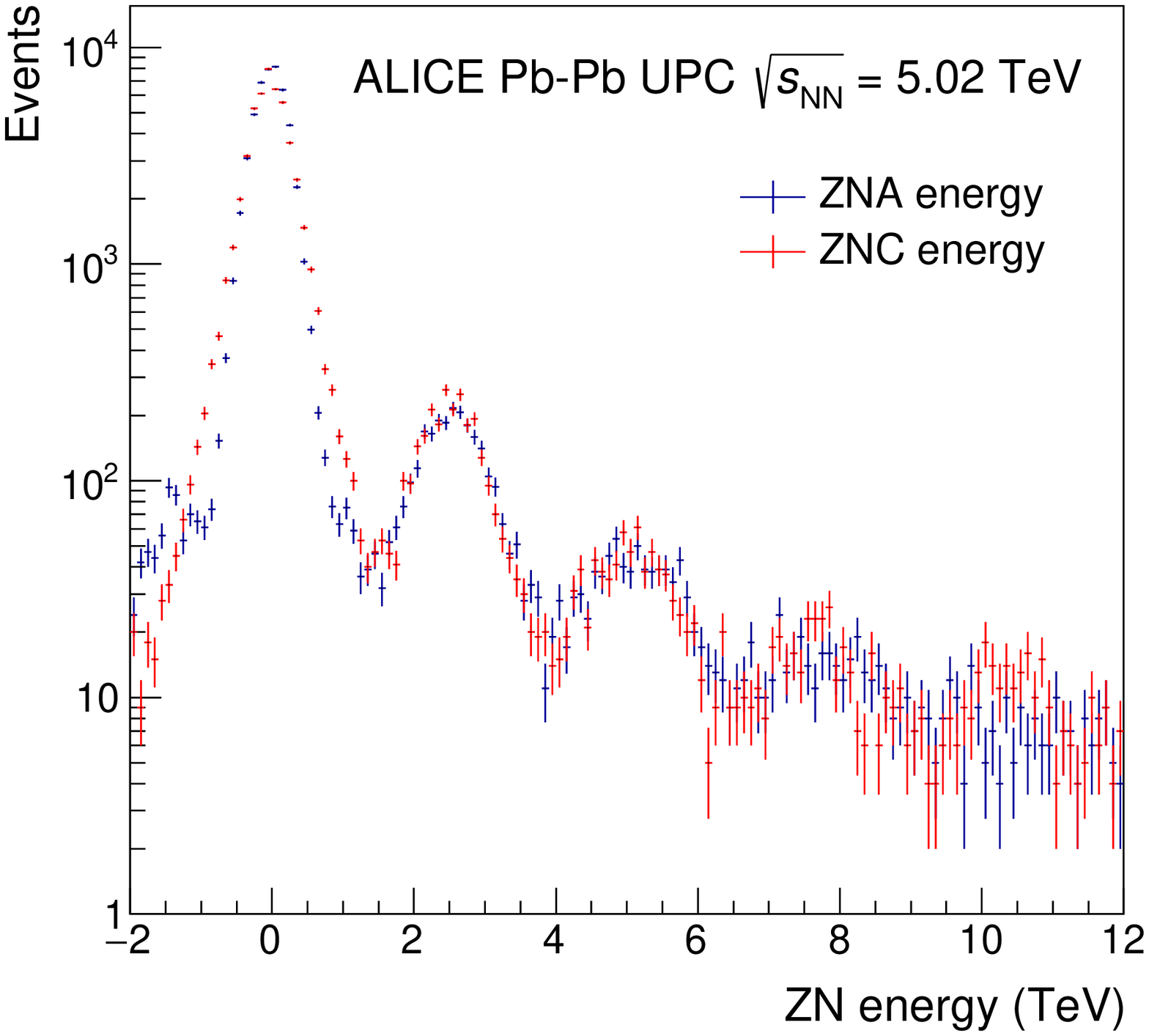}
 \caption{ (Colour online). Correlation between the energy distributions of  the ZNA and ZNC detectors for events selected for the analysis (left).  Energy distribution in each single detector (right). 
  \label{fig:ZDCspectrum}}
\end{figure}

There are two ZDC detectors, ZNA and ZNC, dedicated to the measurement of neutrons at beam rapidity~\cite{ALICE:2012aa}. They are  located at either side of the nominal interaction point at $\pm 112.5$ m along the $z$-axis. These calorimeters determine the arrival time of the particles allowing  beam--beam and beam--gas interactions to be separated. Furthermore, they  have a good efficiency  to detect neutrons with  $|\eta|>8.8$ and have a relative energy resolution of around 20\% for single neutrons, which allows for a clear separation of events with either zero or a  few neutrons at beam rapidities. This is illustrated in Fig.~\ref{fig:ZDCspectrum}, where the concentration of events corresponds to the cases of zero, one, two or more, neutrons detected.

The trigger used to obtain the data sample for the measurements described below is composed of five signals. Four of them veto any activity within the time windows for nominal beam--beam interactions in ADA, ADC, V0A and V0C. In addition, the SPD provides a topological trigger formed by four SPD triggered chips. These chips form two pairs, each pair with two chips falling in compatible azimuthal regions, but in different SPD layers.  The trigger selects events with at least two pairs of chips having an opening angle in azimuth larger than 153 degrees. The reason to request this topology is that the coherently produced $\Rz$ has very small transverse momentum, and thus the two pions from its decay are produced almost back-to-back in azimuth. 

The integrated luminosity is determined using a reference trigger based on the multiplicity of the V0A and V0C detectors. The corresponding cross section is obtained using a Glauber model for hadronic Pb--Pb collisions~\cite{Loizides:2017ack}.  The integrated luminosity for the measurements presented below is 485 mb$^{-1}$ with a relative systematic uncertainty of 5\%.

\section{Analysis procedure
\label{sec:Ana}}
\subsection{Event selection \label{sec:EventSelection}}
Events that fulfil the trigger criteria described above are selected for further analysis if they contain exactly two tracks of opposite electric charge. To ensure a proper measurement, each track is required to have at least 50 space points in the TPC and one associated hit in each layer of the SPD. These SPD hits have to be matched to a triggered readout chip. Furthermore, each track has to have a distance of closest approach to the event interaction vertex of less than 2~cm in the $z$-axis direction and less than $0.0182 + 0.0350/(p^{\rm trk}_{\textrm T})^{1.01}$~cm in the  plane transverse to the beam direction. Here $p^{\rm trk}_{\textrm T}$ denotes the transverse momentum of the track in GeV/$c$.

The energy loss of each reconstructed track is measured in units of the standard deviation ($\sigma_{\pi}$) with respect to  Bethe expectations for a pion passing the TPC. The track pair is accepted if 
\[\left(n^2_{\sigma_{\pi^+}}+n^2_{\sigma_{\pi^-}}\right)<5^2.\] This criterion rejects, in the considered mass range, the contribution from electrons, while there remains a small background from muon pairs which is discussed below.

The four momentum of the track pair is computed under the assumption of each track being a pion. A pair is accepted if its rapidity ($y$), transverse momentum ($p_{\textrm T}$) and mass ($m$) are within $|y|<0.8$, ${p_{\textrm T}<0.2}$~GeV/$c$ and  $0.55<m<1.4$~GeV/$c^2$.

To veto activity in the pseudorapidity range covered by the AD and V0 detectors, their offline signals are studied. The offline reconstruction in these detectors is more precise than the online information, because it uses larger time windows than the trigger electronics and a more refined algorithm to quantify the signal. Events showing a reconstructed signal in any  of ADA, ADC, V0A or V0C are rejected.

\begin{figure}[t!]
\centering
 \includegraphics[width=0.48\textwidth]{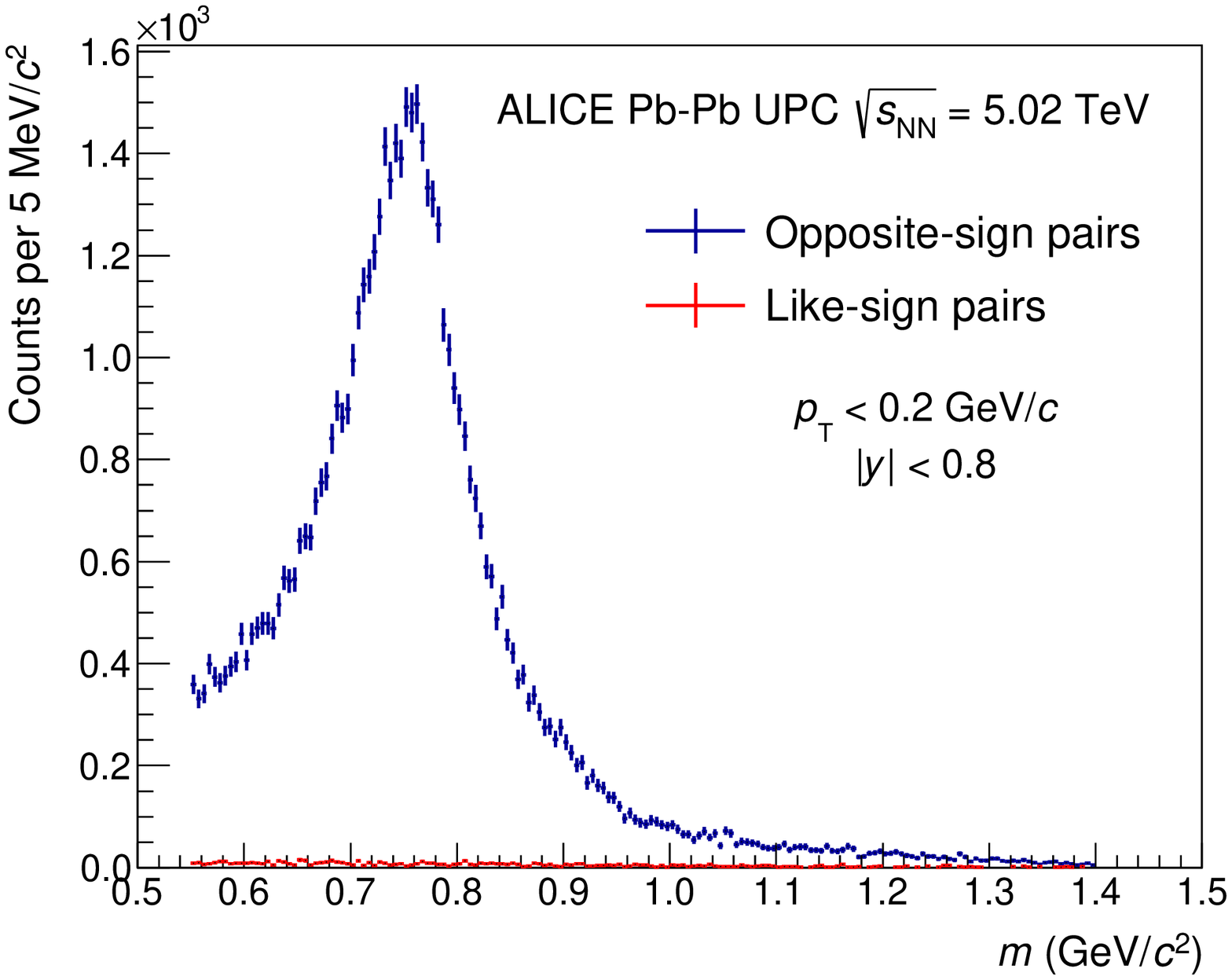} 
  \includegraphics[width=0.46\textwidth]{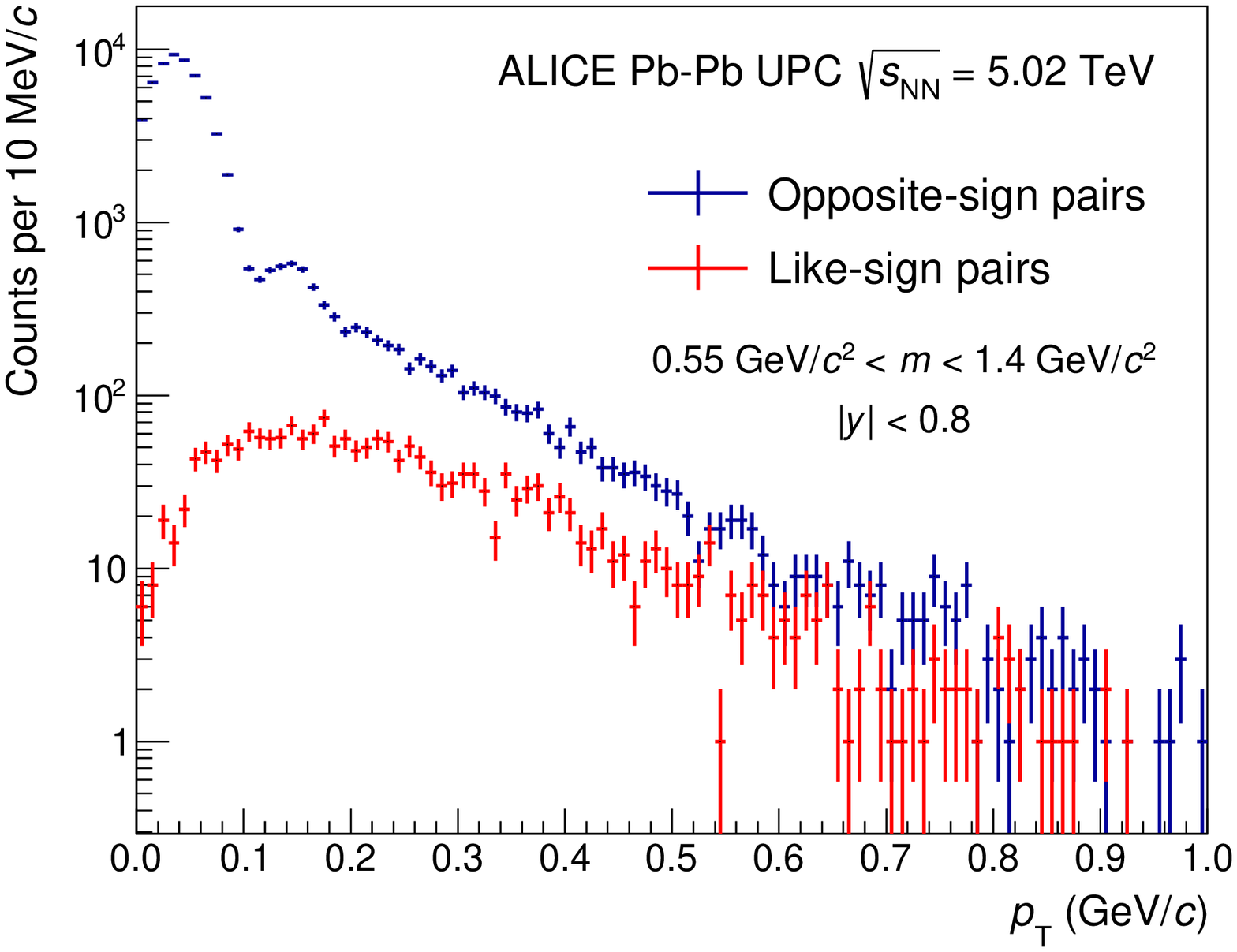} 
 \caption{ (Colour online).   Invariant mass (left) and transverse momentum (right) distributions for opposite-sign (blue) and like-sign (red) pairs.
  \label{fig:RawData}}
\end{figure}

The invariant mass distribution for $p_{\textrm T}<0.2$~GeV/$c$ and transverse momentum distribution for $0.55<m<1.4$~GeV/$c^2$ of the selected track pairs are shown in Fig.~\ref{fig:RawData}. The mass distribution shows the  shape expected from a $\Rz$ spectrum, while  a diffraction dip is clearly seen in the transverse momentum distribution. In total, the signal sample contains almost 57 thousand events which passed all selection criteria. 

The signal sample is further subdivided in forward-neutron classes. The assignment of an event to a class is based on the timing capabilities of the ZNA and ZNC detectors. Events in which the timing of the energy deposition in the calorimeter is consistent within $\pm2$ ns with the neutron having been produced in a beam--beam collision are classified as having a forward neutron in the corresponding calorimeter. 

\subsection{Background subtraction and corrections for experimental effects
\label{sec:background}}

In this section the procedure to determine the corrections used in the measurement is presented. The correction factors are quoted with their corresponding uncertainties, which are discussed in Sec.~\ref{sec:SysUnc} and summarised in Tables~\ref{tab:sys} to~\ref{tab:sysZDC}.

As a tool to quantify some of the remaining background contributions a special sample of events is selected fulfilling all criteria mentioned in Sec.~\ref{sec:EventSelection}, except that both tracks  have the same electric charge. 
The invariant mass and transverse momentum distributions of this sample are shown in Fig.~\ref{fig:RawData}. The distributions of  same-charge pairs are used as an estimation of the amount and shape of the background from events with a measured opposite-charge pair and  other charged tracks outside the acceptance of the detector.  The contribution of same-charge pairs 
is at the level of 1\% and is statistically subtracted from the signal sample.

Another potential background comes from events with two tracks with opposite electric charge and a neutral particle. The main contribution  is expected from three-body decays of the $\omega$ vector meson. Dedicated Monte Carlo (MC) simulations of coherent $\omega$ photoproduction followed by the $\omega\rightarrow\pi^+\pi^-\pi^0$ decay demonstrate that the signal from such $\pi^+\pi^-$ pairs from three-body $\omega$ decays concentrates at lower masses and higher transverse momenta than those considered for the signal-extraction procedure described below. This study, as well as all studies involving MC, use generated MC events, in this case from STARlight, passed through a detailed simulation of the ALICE detector.

The  contribution from $\Rz$ vector mesons produced in incoherent interactions is estimated by fitting a template produced by STARlight of the transverse momentum distribution. The template is fitted in the region of transverse momentum  $0.25<p_{\textrm T}<0.9$~GeV/$c$ to obtain its proper normalisation. The normalised template is used to estimate this contribution for $p_{\textrm T}<0.2$~GeV/$c$. The final yield of $\Rz$ mesons is corrected by subtracting this contribution,  which is $(4\pm0.5)$\%.

The efficiency of the SPD readout chips participating in the trigger is measured with a data-driven approach using a minimum bias trigger. Tracks  selected without  requiring  two hits in the different SPD layers are matched to the readout chips they cross. A chip inefficiency affects each track, and thus each event differently. The efficiency maps obtained from data are incorporated into the Monte Carlo simulation of the signal and applied event-by-event. The overall effect corresponds to a global correction of about $(17\pm1)$\%.

The efficiency of the ZNA and ZNC to detect neutrons is estimated with two different methods. 
In the first method, a sample of MC events generated with the RELDIS program~\cite{Pshenichnov:2001qd,Pshenichnov:2011zz} is used. The other method relies on a simple probabilistic model~\cite{Dmitrieva:2018rpi} applied directly to the raw data. Both methods yield compatible results, namely an efficiency of about $(93\pm1)$\% each for the ZNA and ZNC to detect neutron activity. The propagation of this effect, and the one discussed next, into the value of the measured cross sections is discussed in Sec.~\ref{sec:SysUnc}.

Good events in the 0nXn and XnXn classes are rejected when, in addition to the forward neutrons, other particles are created at large rapidities and leave a signal either in the AD or the V0 detectors. These extra particles come from the different possibilities of dissociation of nuclei, e.g. neutron emission, multi-fragmentation or pion production, and the corresponding cross sections are expected to be large~\cite{Pshenichnov:1999hw}. The amount of good events with neutrons which are lost due to vetoes by AD and V0 is estimated using control triggers. The corrections amount to $(26\pm4)$\% for events with a signal either in ZNA or in ZNC, while it is $(43\pm5)$\% for events with a signal in both ZNA and ZNC. 

Good  events are also rejected if  another interaction creates a signal in one of the veto detectors,  an effect known as pile-up. The main pile-up comes from purely electromagnetic interactions producing a low mass electron-positron pair. The probability of the occurrence of pile-up  is correlated with the average number of inelastic hadronic collisions per bunch crossing  ($\mu$), which  for the data used in this analysis varied from ${\mu=0.0002}$ to ${\mu=0.0015}$. The effect of pile-up is estimated using two different methods. One method uses an event sample obtained with an unbiased trigger based only on the timing of bunches crossing the interaction region. This sample is separated into periods with specific $\mu$ values. The probability of a signal in each of the veto detectors is computed for each value of $\mu$ in  otherwise empty events using the unbiased sample. This probability  exhibits a linear behaviour as a function of $\mu$.  The veto inefficiencies
 are determined by weighting the corresponding veto rejection probabilities
over periods with different $\mu$, taking the luminosity of each period as a weight. The correlation between the online and offline vetoes is taken into account. 
The second method divides the signal sample described in Sec.~\ref{sec:EventSelection} into subsets of events with a specific range of $\mu$ values. Each one of these sub-samples is subjected to the full analysis chain. The final cross sections show a linear dependence on $\mu$. The intercept at $\mu=0$ is taken as the pile-up corrected cross section in this method. The two approaches produce slightly different results. The average of both results is used as the final correction factor of  $(11.1\pm3.8)$\%.

Pile-up also affects the classification on forward-neutron classes. Electromagnetic dissociation processes~\cite{ALICE:2012aa} have a large cross section and produce neutrons at beam rapidities. Using the same unbiased sample as described above, the average pile-up probability 
is measured to be $(3.3\pm0.3)$\% in both ZNA and ZNC.

Finally, the product of the acceptance times efficiency  to measure the coherently produced $\Rz$ vector meson is determined using event samples generated with STARlight. Two different samples are used: one of pure coherent $\Rz$ photoproduction  and  the other produced with a flat mass distribution. Both approaches yield similar correction functions for the invariant mass spectrum. 
The acceptance times efficiency rises smoothly from 15\% to 19\% in the mass range from 0.6~GeV/$c^2$ to 1.2~GeV/$c^2$ and remains constant for larger masses.

\subsection{Signal extraction}
The  invariant mass distribution, corrected by all effects described above and normalised by the luminosity of the sample, is fitted to the sum of a S\"oding formula~\cite{Soding:1965nh} and a term $M$ to account for the contribution of the $\gamma\gamma\to\mu^+\mu^-$ process:
\begin{equation}
	\frac{{\rm d}\sigma}{{\rm d}m\,{\rm d}y} = |A \cdot BW_{\rho}+B|^2+M,
	\label{Soeding}
\end{equation}
where $A$ is the normalisation factor of the $\Rz$ Breit-Wigner ($BW_{\rho}$) function, and $B$ is the non-resonant amplitude. The relativistic Breit-Wigner function of the $\Rz$ vector meson is
\begin{equation}
	BW_{\rho} = \frac{\sqrt{m \cdot m_{\rho^0} \cdot \Gamma(m)}}
                  {m^2- m^2_{\rho^0} + im_{\rho^0} \cdot \Gamma(m)},
	\label{SoedingBW}
\end{equation}
where $m_{\Rz}$ is the pole mass of the $\Rz$ vector meson. The mass-dependent width $\Gamma(m)$ is given by
\begin{equation}
	\Gamma(m) = \Gamma(m_{\rho^0}) \cdot \frac{m_{\rho^0}}{m} \cdot 
         \left( \frac{m^2-4m^2_{\pi}}{m^2_{\rho^0}-m^2_{\pi}} \right)^{3/2},
	\label{SoedingWidth}
\end{equation}
with $\Gamma(m_{\Rz})$ the width of the $\Rz$ vector meson and $m_{\pi}$  the mass of the pion~\cite{Jackson:1964zd}.

Instead of Eq.~(\ref{Soeding}), one could also consider an extended model that includes a term for the production of $\omega$ vector mesons as done recently by STAR~\cite{Adamczyk:2017vfu}.  The use of such a model would not affect the results presented here for the $\Rz$ vector meson, but, unfortunately, the size of the current data sample does not allow for the extraction of the parameters related to $\omega$ production.

 The invariant mass dependence of the $\gamma\gamma\to\mu^+\mu^-$ process is obtained by a sample of events from STARlight which are passed through a detailed simulation of the ALICE detector and scaled using the correction factors obtained for the $\Rz$ case. The normalisation is fixed to the cross section predicted by STARlight, because this MC correctly describes  the cross section of the $\gamma\gamma\to {\rm e}^+{\rm e}^-$ process in the $\Rz$ mass range in our previous measurement~\cite{Kryshen:2019jnz}.

\begin{figure}[t!]
\centering
 \includegraphics[width=0.48\textwidth]{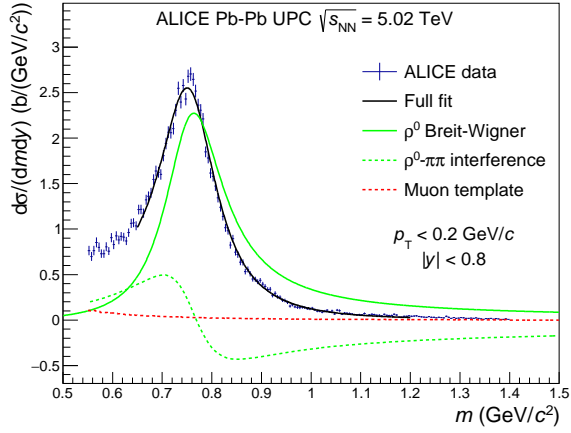} 
 \caption{ (Colour online).   
 Invariant mass distribution of pion pairs with the different components of the fit represented by lines. See text for details.
  \label{fig:Fit}}
\end{figure}

An example of this fit  is shown in Fig.~\ref{fig:Fit}.  A clear signal for the $\Rz$  vector meson is visible. The contribution from the $\gamma\gamma\to\mu^+\mu^-$ process is small.
The values found for the pole mass and width of the $\Rz$ are $769.5 \pm 1.2$ (stat.) $\pm$ 2.0 (syst.) MeV/$c^2$ and 156~$\pm$~2 (stat.) $\pm$ 3 (syst.) MeV/$c^2$, respectively. (The estimation of the systematic uncertainty is described below.) These values are consistent with those reported  by the PDG~\cite{Tanabashi:2018oca}. These parameters are then fixed to their PDG values when extracting the coherent $\Rz$ yield. 

Following the standard convention, the cross section for coherent  production of $\Rz$ vector mesons in UPC is extracted by integrating the BW$_{\rho}$ component of the fit in the invariant mass range from $2 m_{\pi}$ to $m_{\Rz}+ 5\Gamma(m_{\rho^0})$.
Measurements are reported for the following ranges of rapidity: $|y|<0.2$, $0.2<|y|<0.45$, and $0.45<|y|<0.8$. The ranges are chosen to have approximately the same number of pion pairs and to have a  number of pairs large enough to allow for a meaningful measurement of the cross sections for the different forward-neutron classes.

\begin{figure}[t!]
\centering
  \includegraphics[width=0.48\textwidth]{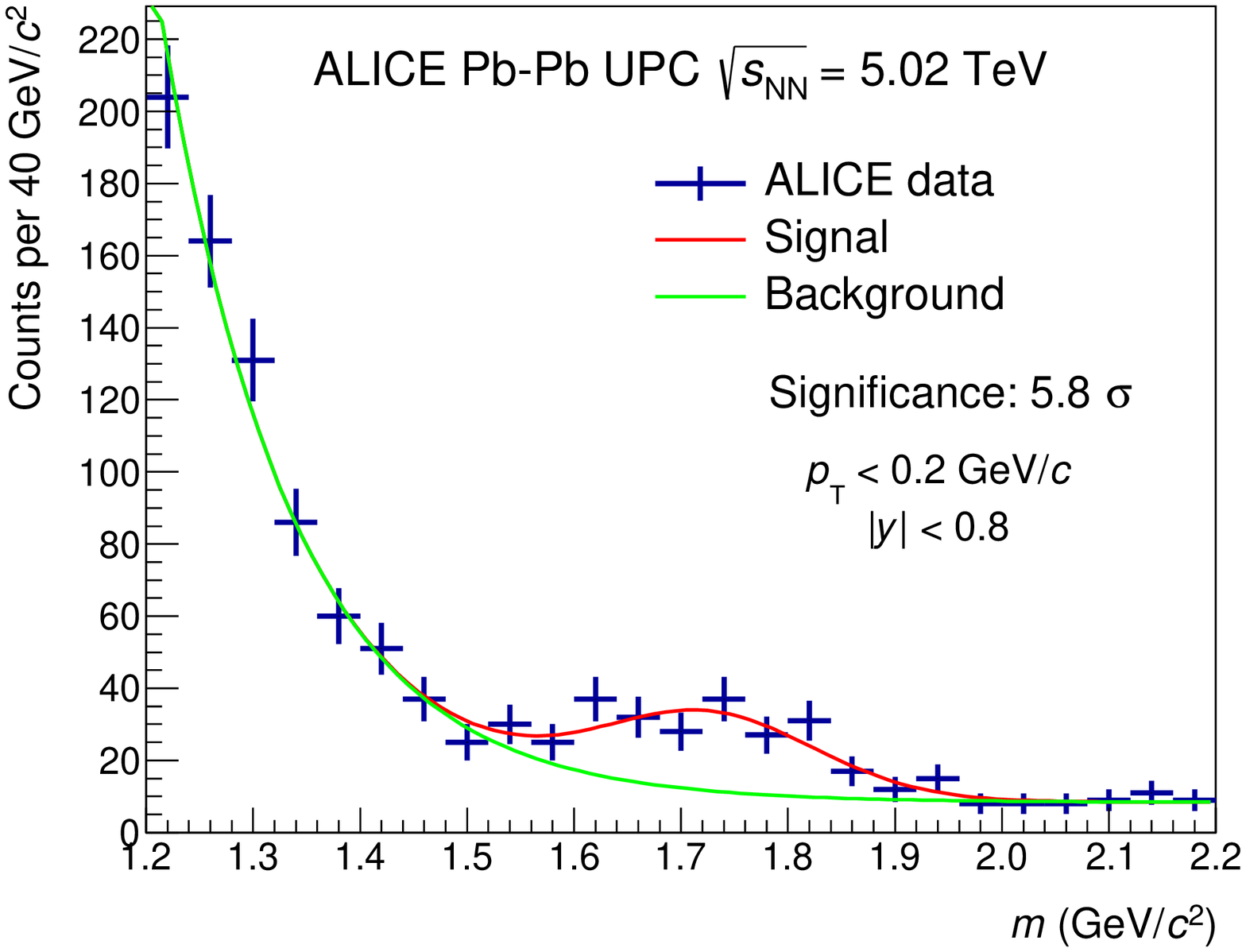}
  \includegraphics[width=0.48\textwidth]{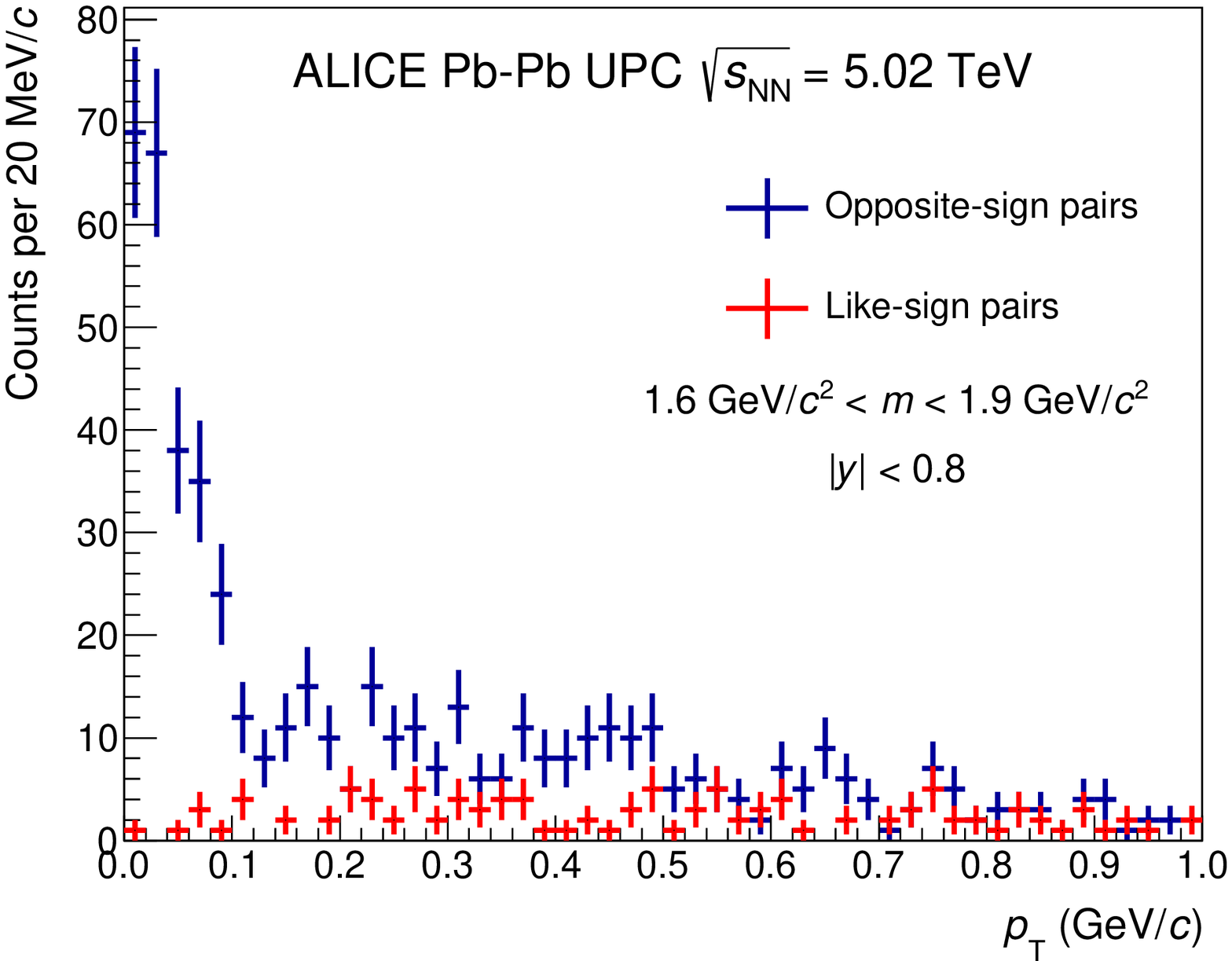}  
 \caption{ (Colour online).   Invariant mass (left) and transverse momentum (right) distributions of pion pairs  at large invariant masses. The lines correspond to the fit components described in the text. The symbols depict the signal and like-sign background  distributions.
  \label{fig:HM}}
\end{figure}

\subsection{Signal extraction at large invariant masses}
The uncorrected invariant mass distribution for $m>1.2$~GeV/$c^2$  is shown in  Fig.~\ref{fig:HM}, which also shows the distribution of transverse momentum for pion pairs in the mass range from 1.6~GeV/$c^2$ to 1.9~GeV/$c^2$.  The latter distribution  peaks at small $p_{\rm T}$, as expected from coherent production. The contribution of like-sign pion pairs is very small in this region. The invariant mass distribution is fitted with the same model as used by  the STAR Collaboration~\cite{Klein:2016dtn},

\begin{equation}
	\frac{{\rm d}N_{\pi\pi}}{{\rm d}m} = P_1 \cdot \exp{(-P_2 \cdot (m-1.2\, {\rm GeV/c^2}))} + P_3 + P_4 \cdot \exp{(-(m-M_x)^2/\Gamma_x^2)},
	\label{HighMassFitFormula}
\end{equation}
where $N_{\pi\pi}$ is the number of pion pairs, $P_i$ are parameters describing the background and the normalisation of the Gaussian part, and $M_x$ ($\Gamma_x$) represent the mass (width) of a potential resonance. As mentioned previously, the acceptance times efficiency correction factor in this mass range is fairly flat, so the uncorrected spectrum is a good approximation of the real one.

The fit to the invariant mass distribution shown in Fig.~\ref{fig:HM} yields a $\chi^2/{\rm d.o.f.}$ of $13/19$. A fit without the contribution of the Gaussian component yields a $\chi^2/{\rm d.o.f.}$ of $63/22$. This rejects the hypothesis that the Gaussian is absent at a significance level of $4.5$ standard deviations. Estimating the significance, $s$, of the Gaussian component by $S/\sqrt{S+2B}$ yields $s = 5.8$, with the signal $S=140\pm16$ and the background $B=222\pm20$, both counted in the mass range $(M_x-2\Gamma_x,M_x+2\Gamma_x)$. 

\subsection{Systematic uncertainties
\label{sec:SysUnc}}
The fit to extract the $\Rz$ contribution, see Eq.~(\ref{Soeding}), is repeated choosing random combinations of the lower and upper limits of the fit range, as well as of the bin width. The lower and upper  limits are varied in the ranges (0.6--0.65)~GeV/$c^2$ and (1.0--1.4)~GeV/$c^2$, respectively, while the bin widths are varied from 0.05 to 0.2~GeV/$c^2$. The results reported below, as well as the above quoted values for the pole mass and width of the $\Rz$, are the average of the values obtained in these fits, while the RMS provides the systematic uncertainty, which varies from 0.4\% to 5.9\%, the largest values corresponding to the XnXn sample. The statistical uncertainty is taken as the average of the statistical uncertainty of each one of the fits. This uncertainty is uncorrelated across rapidity and forward-neutron classes.
The fit procedure is performed using both a $\chi^2$ approach and a binned extended log-likelihood. The results from both methods are consistent.

A Ross-Stodolsky function~\cite{Ross:1965qa} is used as an alternative model. This model yields  cross sections  larger by 3.5\% than those obtained from the S\"oding model. A test using random generated data with a S\"oding model fitted with the Ross-Stodolsky function and vice versa was performed. In both cases a similar difference of around 3.0\% was found. As the underlying distribution is not known the 3.5\% difference observed in data is considered as a systematic uncertainty.

The uncertainty on the track selection is estimated by changing the track selection criteria within reasonable values and repeating the full analysis. The uncertainty corresponds to the full variation of the results and amounts to $\pm1.5$\%. The uncertainty on the matching of TPC and ITS tracks is obtained by comparing the behaviour of real and simulated data under different detector conditions; it amounts to $\pm4$\%.

The uncertainty on the acceptance and efficiency to reconstruct the $\Rz$ vector meson is estimated from the full variation of the results when using the two different MC samples discussed above, namely a flat mass distribution or that of a $\Rz$ meson sample. It amounts to $\pm1$\%. 

The uncertainty on the normalisation of the  template for the $\gamma\gamma\to\mu^+\mu^-$ process is estimated as follows. The statistical uncertainty of the $\gamma\gamma\to {\rm e}^+{\rm e}^-$ cross section  in our previous measurement~\cite{Kryshen:2019jnz} is around 10\% and within this precision it agrees with the prediction from STARlight, validating the use of this MC in this mass range. Changing the normalisation of the $\gamma\gamma\to\mu^+\mu^-$ template  in the fit by $\pm10$\%, produces a $\pm0.3$\% systematic uncertainty on the extracted $\Rz$ cross section.

The fit to extract the incoherent contribution is repeated using different lower and upper limits, as well as bin widths. The respective ranges in transverse momenta are (0.25--0.4)~GeV/$c$, (0.6--0.9)~GeV/$c$, and (0.06--0.18)~GeV/$c$. These variations produce a 0.5\% systematic uncertainty. 

The uncertainty associated to the determination of the  trigger efficiency of the SPD chips is obtained by changing the requirements on the events used for this data-driven method. Variations include the running conditions, the maximum amount of activity allowed in the event, and the definition of tracks accepted in the efficiency computation. This uncertainty amounts to 1\%.

The uncertainty on the pile-up correction from the difference of the two procedures described above is $\pm 3.8$\% for the $\Rz$ cross section. The systematic uncertainty due to pile-up contamination affecting the classification on the forward-neutron classes is discussed below.

Cross sections obtained in  positive and negative rapidity ranges agree within statistical uncertainties, as expected by the symmetry of the process. Similarly,  cross sections for the 0nXn  class with neutrons at positive rapidities are compatible within statistical uncertainties with those with the neutrons at negative rapidities. 

Except the first, all other sources of systematic uncertainty discussed above are correlated across different rapidity intervals and forward-neutron classes. They are summarised in Table~\ref{tab:sys}. The total uncertainty is obtained by adding in quadrature the individual contributions.

The uncertainty on the correction for good 0nXn and XnXn accompanied by particle production leaving a signal in the AD and V0 and being rejected due to the vetoes imposed in these detectors are estimated by varying the selection criteria in the control samples as well as by modifying the  pile-up probability in these samples within their uncertainties. The uncertainty on the correction factors amounts to 4\% and 5\% for the 0nXn and XnXn cases, respectively. The effect of these uncertainties on the final cross sections is reported in Table~\ref{tab:sysEMD}. There is an effect in the 0n0n cross section due to the migrations among neutron classes discussed next.

The cross sections for the different forward-neutron classes have another  uncertainty related to  migrations across classes. It is estimated by propagating the uncertainty in the pile-up and efficiency factors in  ZNA and ZNC. The uncertainty in the efficiency is obtained from the comparison between both models used to estimate it (see Sec.~\ref{sec:background}) and amounts to 1\% for both ZNA and ZNC. The uncertainty in the pile-up in ZDC originates from the statistical uncertainty of the different samples of unbiased events for each $\mu$ value  and amounts to 0.3\%. The effect of 
 these uncertainties on the cross sections in forward-neutron classes is summarised in Table~\ref{tab:sysZDC}.
These uncertainties only move events from one class to another, meaning that some of the uncertainties are anti-correlated among the classes. Note  that the 0nXn cross section is particularly sensitive to the pile-up uncertainty. This is due to the large difference in the values of the 0n0n and 0nXn cross sections which, in the case of pile-up, produces sizeable migrations into the 0nXn class.

\begin{table}[t!]
\centering
\caption{Summary of the systematic uncertainties.  See text for details.}
\begin{tabular}{lr}
\hline
 Source & Uncertainty   \\
 \hline
Variations to the fit procedure & 0.4--5.9 \% \\
Ross-Stodolsky fit model & +3.5\%  \\ 
Track selection & $\pm1.5$\% \\
Track matching & $\pm4.0$\%  \\
Acceptance and efficiency & $\pm1.0$\%  \\ 
Muon background ($\gamma\gamma\to\mu^+\mu^-$) & $\pm0.3$\% \\ 
Incoherent contribution& $\pm0.5$\% \\ 
Trigger efficiency of SPD chips & $\pm 1.0$\% \\
Pile-up & $\pm 3.8$\%\\ 
Luminosity & $\pm5.0$\%  \\
\hline
Total & $^{+({\textrm 8.5-10.3})}_{-({\textrm 7.8-9.7})}$ \% \\
\hline
\end{tabular}
\label {tab:sys}
\end{table}

\begin{table}
\centering
\caption{Summary of the systematic uncertainties on the cross sections related to the correction factors to account for the events with neutrons which are vetoed by the AD or V0 detectors. See text for details. The numbers correspond to the  variations of the cross sections in per cent.}
{\renewcommand{\arraystretch}{1.5}
\begin{tabular}{lrrrr} 
\hline
Source        & No forward-neutron selection  & 0n0n   & 0nXn &XnXn    \\ 
\hline
Signal either in ZNA or in ZNC  & $^{-1.0}_{+1.1}$ & $\pm 0.1$ & $^{-6.6}_{+7.3}$ & $^{+0.6}_{-0.7}$  \\
Signal in both ZNA and ZNC & $^{-0.3}_{+0.4}$ & $\pm 0.7$ & $^{+0.3}_{-0.4}$ & $^{-8.9}_{+10.6}$   \\ 
\hline
\end{tabular}
}
\label {tab:sysEMD}
\end{table}

\begin{table}
\centering
\caption{Summary of the systematic uncertainties related to the forward-neutron class selection. The percentile variation of the cross sections is shown. See text for details.}
{\renewcommand{\arraystretch}{1.3}
\begin{tabular}{lrrr} 
\hline
Source          & 0n0n   & 0nXn &XnXn    \\ 
\hline
& \multicolumn{3}{c}{$|y|<0.2$}\\
\hline
ZDC efficiency     & $\mp0.1$      &      $\pm0.6$  & $\pm2.2$    \\ 
ZDC pile-up      & $\mp 0.7$   & $^{+5.4}_{-4.8}$     & $\pm1.4$   \\ 
\hline
&\multicolumn{3}{c}{$0.2<|y|<0.45$}\\
\hline
ZDC efficiency     & $\mp0.1$      &      $\pm0.5$  & $\pm2.2$    \\ 
ZDC pile-up      & $\mp0.7$   & $^{+5.6}_{-5.0}$     & $\pm1.4$                          \\ 
\hline
&\multicolumn{3}{c}{$0.45<|y|<0.8$}\\
\hline
ZDC efficiency     & $\mp0.1$      &      $\pm0.5$  & $\pm2.2$    \\ 
ZDC pile-up      & $\mp0.7$   & $^{+5.5}_{-4.9}$     & $\pm1.3$                          \\ 
\hline
\end{tabular}
}
\label{tab:sysZDC}
\end{table}

\section{Results
\label{sec:Results}}
\subsection{Coherent photoproduction of $\rho^0$ vector mesons}

\begin{figure}[t!]
\centering
 \includegraphics[width=0.48\textwidth]{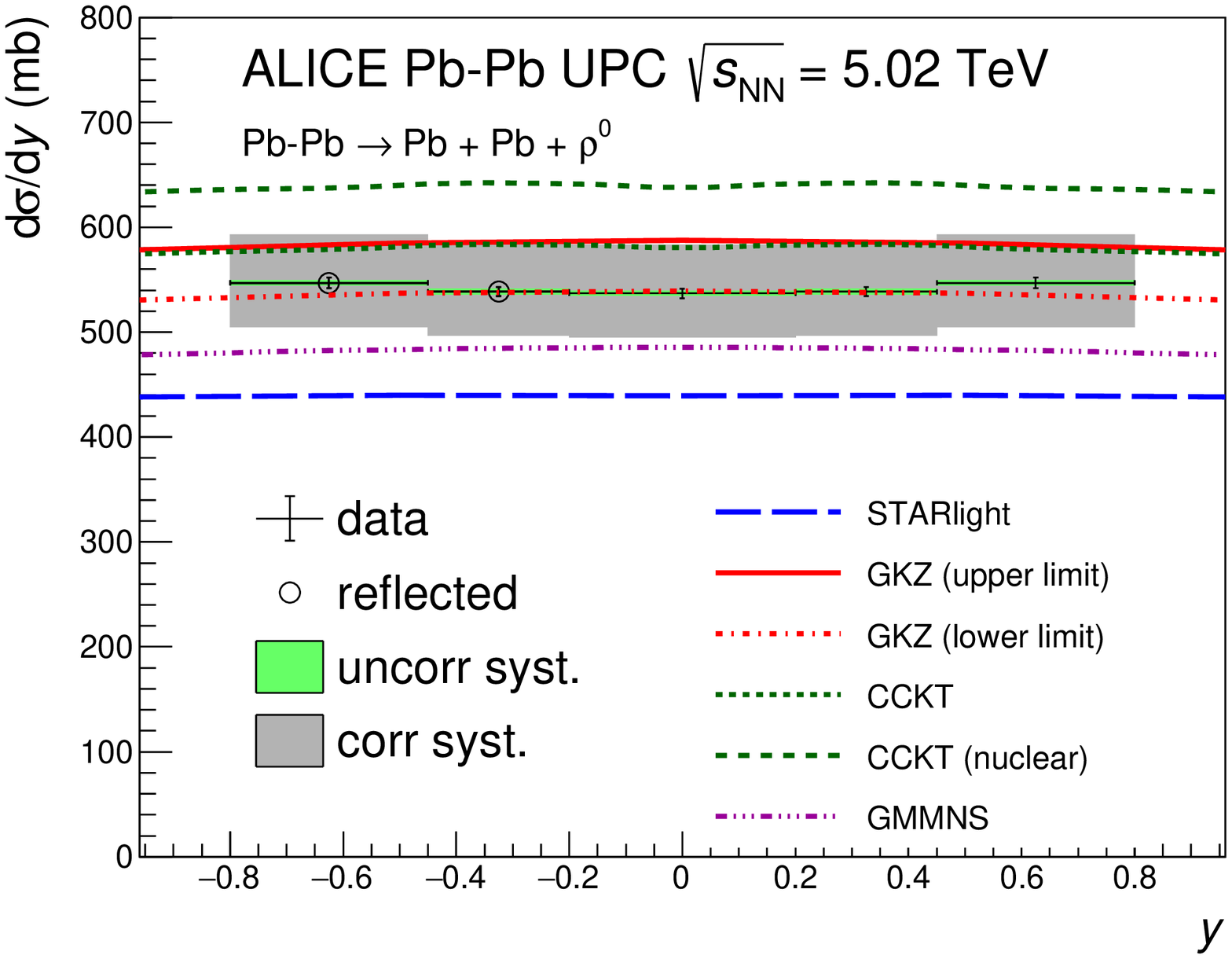} 
 \includegraphics[width=0.48\textwidth]{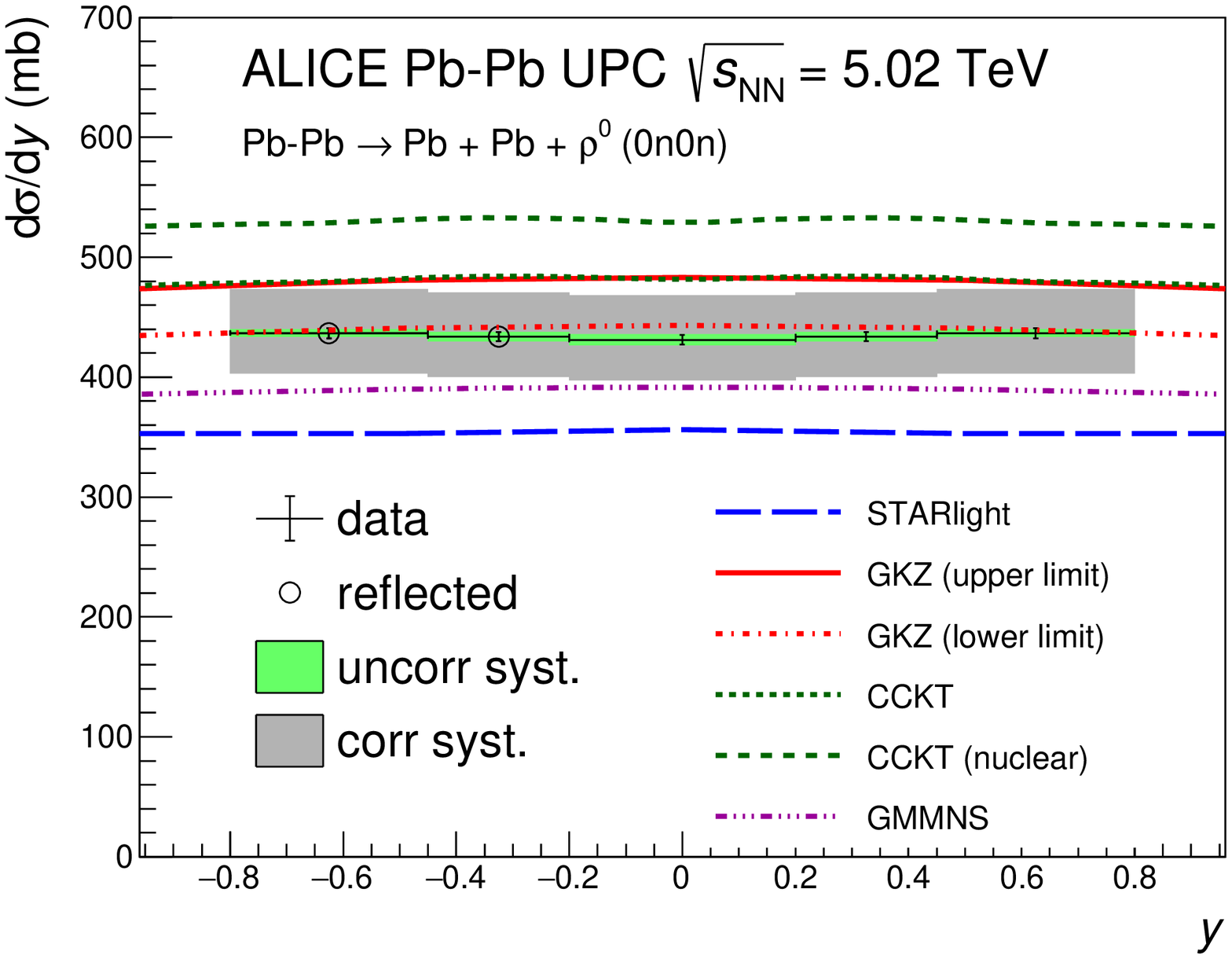} \\
 \includegraphics[width=0.48\textwidth]{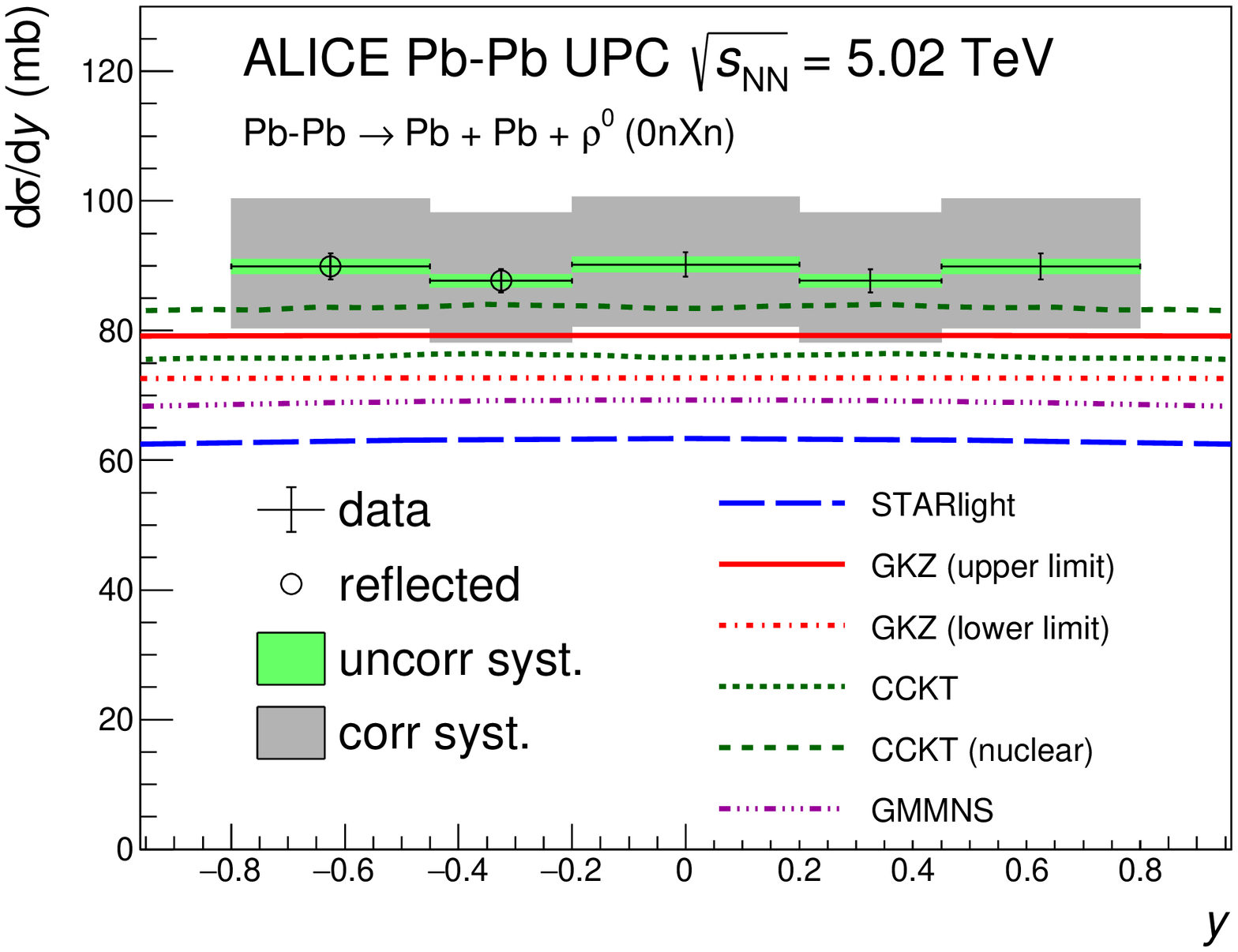} 
 \includegraphics[width=0.48\textwidth]{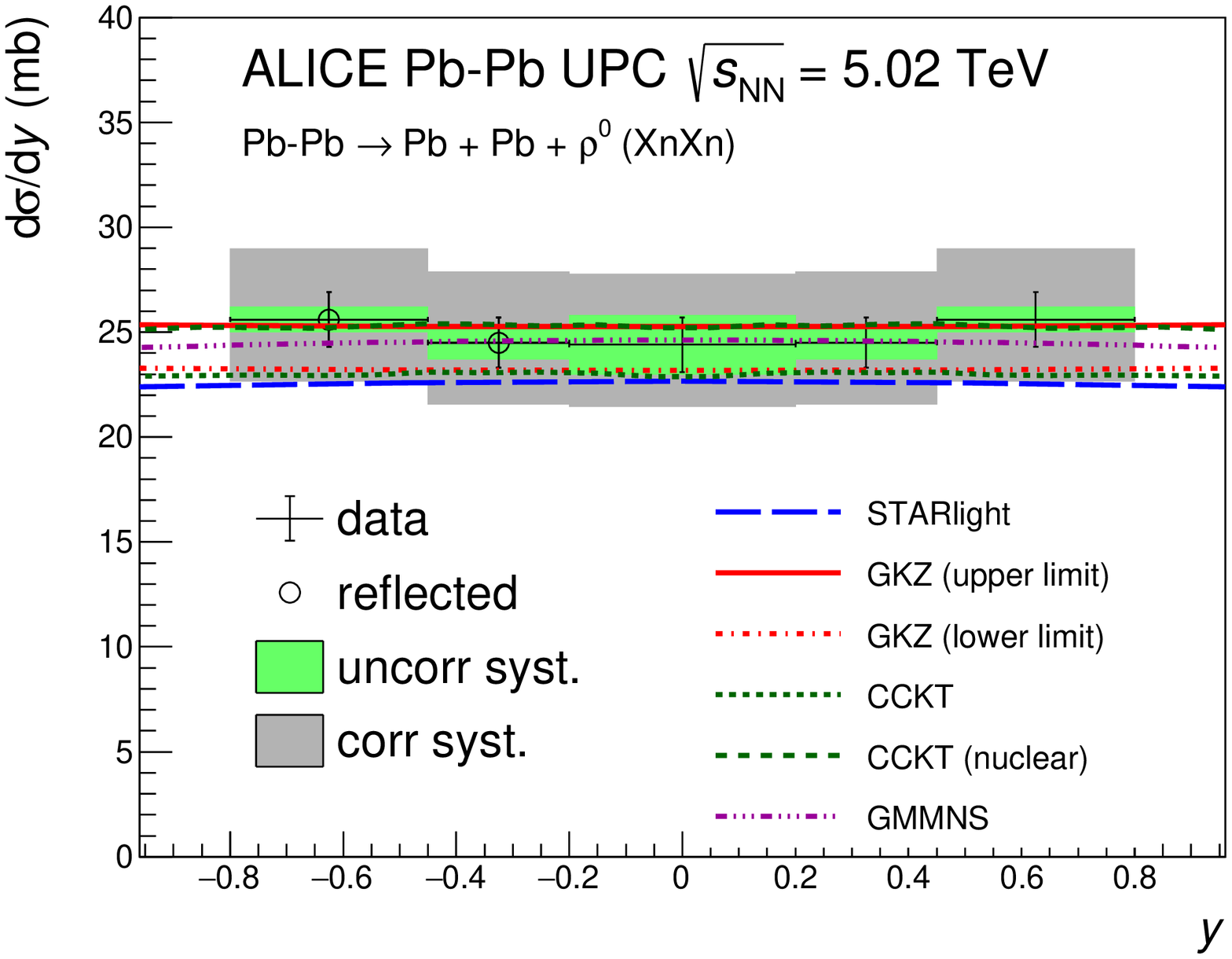} \\
 \caption{ (Colour online).   Cross section for the coherent photoproduction of $\Rz$ vector mesons in Pb--Pb UPC as a function of  rapidity for no forward-neutron selection (top left), and for the 0n0n (top right), 0nXn (bottom left) and XnXn (bottom right) classes. The lines show the predictions of the different models described in the text.
  \label{fig:XS}}
\end{figure}

\begin{table}
\centering
\caption{Numerical values of the cross section for the coherent photoproduction of $\Rz$ vector mesons in Pb--Pb UPC at $ \sqrt{s_{\mathrm{NN}}}= 5.02 $ TeV. The systematic uncertainties are obtained by adding in quadrature the contributions listed in Tables~\ref{tab:sys} to~\ref{tab:sysZDC}.}
{\renewcommand{\arraystretch}{1.3}
\begin{tabular}{lrrr}
\hline 
 No forward-neutron selection  &   Cross section (mb)  & stat. (mb) & syst. (mb) \\
\hline 
$|y|<0.2$ & 537.0 & 4.6 & $^{+46.1}_{-42.0}$ \\
$0.2<|y|<0.45$ & 538.6 & 4.4 & $^{+46.2}_{-42.1}$ \\
$0.45<|y|<0.8$ & 547.0 & 4.9 & $^{+46.9}_{-42.8}$ \\
\hline 
0n0n & & \\ 
\hline
$|y|<0.2$ & 431.1 & 4.0 & $^{+36.8}_{-33.6}$ \\
$0.2<|y|<0.45$ & 433.8 & 3.8 & $^{+37.0}_{-33.8}$ \\
$0.45<|y|<0.8$ & 436.7 & 4.2 & $^{+37.3}_{-34.0}$ \\
\hline 
0nXn & & \\ 
\hline
$|y|<0.2$ & 90.2 & 1.9 & $^{+10.5}_{-9.5}$ \\
$0.2<|y|<0.45$ & 87.7 & 1.8 & $^{+10.2}_{-9.3}$ \\
$0.45<|y|<0.8$ & 89.9 & 2.0 & $^{+10.4}_{-9.5}$ \\
\hline 
XnXn & & \\ 
\hline
$|y|<0.2$ & 24.4 & 1.3 & $^{+3.4}_{-2.9}$ \\
$0.2<|y|<0.45$ & 24.5 & 1.2 & $^{+3.4}_{-3.0}$ \\
$0.45<|y|<0.8$ & 25.6 & 1.3 & $^{+3.5}_{-3.1}$ \\
\hline \end{tabular}
}
\label {tab:xs}
\end{table}

Figure~\ref{fig:XS} shows the cross section for the coherent photoproduction of $\Rz$ vector mesons in Pb--Pb UPC as a function of  rapidity. The measurements are performed for ranges in the absolute value of rapidity. For display purposes, the measurements are shown in Fig.~\ref{fig:XS} at positive rapidities and reflected into negative rapidities. The cross sections are reported numerically in Table~\ref{tab:xs}. Data are compared with the following models:
\begin{itemize}
\item[]{\bf STARlight}. This model is based on a phenomenological description of the exclusive production of $\Rz$ vector mesons off nucleons, the optical theorem, and a Glauber-like eikonal formalism, neglecting the elastic part of the elementary $\Rz$--nucleon cross section, to describe nuclear effects~\cite{Klein:1999qj,Klein:2016yzr}.
\item[]{\bf GKZ}. These predictions by Guzey, Kryshen and Zhalov (GKZ) are based on a modified vector-dominance model, in which the hadronic fluctuations of the photon interact with the nucleons in the nucleus according to the Gribov-Glauber model of nuclear shadowing. The model is introduced in~\cite{Frankfurt:2015cwa}, while the predictions for 
Pb--Pb UPC at $\snn = 5.02$~TeV are presented in~\cite{Guzey:2016piu}. In the figures the variations of the prediction on the uncertainty of theory parameters are shown as upper and lower limit of the model; see ~\cite{Frankfurt:2015cwa} for details.
\item[]{\bf GMMNS}. This model by Goncalves, Machado, Morerira, Navarra and dos Santos (GMMNS)~\cite{Goncalves:2017wgg} is based on the Iancu-Itakura-Munier (IIM)~\cite{Iancu:2003ge} implementation of gluon saturation within the colour-dipole model coupled to a boosted-Gaussian description of the wave function of the vector meson.
\item[]{\bf CCKT}. This model by Cepila, Contreras, Krelina and Tapia (CCKT) is based on the colour-dipole model with the structure of the nucleon in the transverse plane described by so-called hot spots, regions of high gluonic density, whose number increases with increasing energy~\cite{Cepila:2016uku,Cepila:2018zky}. The nuclear effects are implemented along the ideas of the Glauber model proposed in~\cite{Armesto:2002ny}. To highlight the effect of sub-nucleon structure, two versions of the model are presented: one without hot spots (marked as {\em nuclear} in the figures) and  one including the hot-spot structure.
\end{itemize}

The modification of the photon flux due to the emission of the forward neutrons is carried out in the first three models as proposed in~\cite{Baltz:2002pp}. The fourth model uses the $\textbf{n$\mathbf{_O^O}$n}$ afterburner described in~\cite{Broz:2019kpl}.

Figure~\ref{fig:XS}
shows that the lower limit of the GKZ model gives a good description of the 0n0n cross section and underestimates a little bit the 0nXn and XnXn cross sections while the upper limit of the same model overestimates the 0n0n, slightly underestimates the 0nXn and describes the XnXn cross sections.  The STARlight predictions underestimate all the cross section at around the 2 sigma level, except XnXn where the difference is smaller. The behaviour of the CCKT model based on hot spots is quite similar to the upper limit of GKZ; the CCKT (nuclear) variant of this model is some 10\% larger than the  predictions of the CCKT model with hot spots. Finally, the GMMNS model predicts cross sections larger than  STARlight, but still underestimating the measurements  except in the XnXn class.
Taking into account the spread of the models and the uncertainties of data the agreement between the models and the measurement appears in most cases satisfactory, particularly for the predictions of the GKZ model. This overall description of data by models suggests  that the method to obtain the individual photonuclear contributions to the coherent production of $\Rz$ using forward-neutron classes~\cite{Baltz:2002pp,Guzey:2013jaa}  may be applied to the data, specially once the uncertainties in the measurements are reduced and the spread on the theoretical predictions is better understood.

\subsection{Contributions from continuum production}
The $|B/A|$ ratio, see Eq.~(\ref{Soeding}), quantifies the contribution of the continuum in relation to the resonance production cross section.  The value found at midrapidity for no forward-neutron selection is $0.57\pm 0.01\ ({\rm stat.)}\pm 0.02\ ({\rm syst.)}$ $({\rm GeV}/c^2)^{-\frac{1}{2}}$, where it has been checked that most of the effects cancel in the ratio and the only remaining contribution to the systematic uncertainty are the variations in the fit procedure. The measured value can be compared with that found for the same process at $\snn=2.76$~TeV: $0.50\pm0.04\ ({\rm stat.)}^{+0.10}_{-0.04}\ ({\rm syst.)}$ $({\rm GeV}/c^2)^{-\frac{1}{2}}$~\cite{Adam:2015gsa}.  
Within the current systematic uncertainties, the ratio can be taken as constant both as a function of rapidity and for the different forward-neutron classes. 
Nonetheless data seems to indicate a small decrease of the ratio with rapidity for the no forward selection case: $|B/A| = 0.56\pm 0.01\ ({\rm stat.)}\pm 0.02\ ({\rm syst.)}$ $({\rm GeV}/c^2)^{-\frac{1}{2}}$ and $|B/A| = 0.52\pm 0.01\ ({\rm stat.)}\pm 0.01\ ({\rm syst.)}$ $({\rm GeV}/c^2)^{-\frac{1}{2}}$ for the $0.2<|y|<0.45$ and $0.45<|y|<0.8$ intervals, respectively. It would be interesting if such a trend is observed with the large data sample and the improved precision, expected from the LHC Run 3 and 4~\cite{Citron:2018lsq}.

The corresponding ratio in coherent Au--Au UPC measured by
STAR at $\snn = 200$~GeV is  $0.79\pm0.01\ ({\rm stat.)}\pm0.08\ ({\rm syst.)}$ $({\rm GeV}/c^2)^{-\frac{1}{2}}$~\cite{Adamczyk:2017vfu}. These results for production off heavy nuclear targets, can be compared with those from exclusive $\Rz$ photoproduction off protons. Note that value of $|B/A|$ might depend on the range in $|t|$ selected to perform the measurement, where $t$ is the square of the four momentum transfer at the target vertex. The CMS Collaboration measured $0.50\pm0.06\ ({\rm stat.)}$ $({\rm GeV}/c^2)^{-\frac{1}{2}}$ in p--Pb UPC at $\snn=5.02$~TeV~\cite{Sirunyan:2019nog} for $|t|<0.5$~GeV$^2$. The ZEUS Collaboration, using a sample of positron--proton collisions at a centre-of-mass energy of 300 GeV,  reports $0.67\pm0.02\ ({\rm stat.)}\pm0.04\ ({\rm syst.)}$ $({\rm GeV}/c^2)^{-\frac{1}{2}}$ for their full analysed sample, and $\approx 0.8$ $({\rm GeV}/c^2)^{-\frac{1}{2}}$ for $t$ values similar to those of  coherent $\Rz$ production in Pb--Pb UPC~\cite{Breitweg:1997ed}.
Overall, the ratio of the continuum to the resonance production of $\pi^+\pi^-$ pairs seems to be sensitive to both the kinematics of the interaction and the type of target, but no clear picture has yet emerged.

\subsection{Observation of a resonance-like structure}
As shown in Fig.~\ref{fig:HM}, there seems to be a resonance-like structure in the region $m>1.2$~GeV/$c^2$. The model of Eq.~(\ref{HighMassFitFormula}) yields  a mass of $(1725\pm17)$~MeV/$c^2$ and width $(143\pm21)$~MeV/$c^2$, where the quoted uncertainties
correspond to statistical fluctuations only. As shown in the same figure, this resonance-like object has very low transverse momentum as expected from a coherent-production process. 

Such an object is also seen by  the  STAR Collaboration~\cite{Klein:2016dtn} albeit at a slightly lower mass of 1.65~GeV/$c^2$, but with a similar width. ZEUS  reports a peak around 1.8~GeV/$c^2$ for exclusive electroproduction of $\pi^+\pi^-$ pairs~\cite{Abramowicz:2011pk}. More recently, H1 reports a peak at 1.6~GeV/$c^2$ in the exclusive photoproduction of the $\Rz$ meson~\cite{h1prelim}.
As suggested in~\cite{Klein:2016dtn}, this resonance is also compatible with the $\rho_3(1690)$ listed in the PDG, which has a total angular momentum $J=3$~\cite{Tanabashi:2018oca}. 

The large data samples expected in Run 3 and Run 4 at the LHC~\cite{Citron:2018lsq} may help to shed light on the origin and structure of this object.

\section{Summary and outlook
\label{sec:Summary}}
The rapidity dependence of the coherent $\Rz$ vector meson production cross section in Pb--Pb UPC at $\snn = 5.02$~TeV has been presented. In each rapidity range, the cross section is measured for different classes of events defined by the presence of neutrons at beam rapidities. The cross sections  are compared with the main available models of this process.
The measurements of coherent $\Rz$  photoproduction  are in good agreement both with models following the parton-based colour-dipole approach and with the framework of Gribov-Glauber shadowing based on hadronic degrees of freedom. The models~\cite{Baltz:2002pp,Broz:2019kpl} of electromagnetic nuclear dissociation accompanying vector meson photoproduction provide a satisfactory description of the measured  cross sections for different neutron emission classes. This observation suggests that the method proposed in~\cite{Guzey:2013jaa} to decouple the low-photon-energy from the high-photon-energy contribution to the UPC cross section using neutron-differential measurements might also be  applicable at forward rapidities, which is specially important in view of the expected data samples to be recorded at the LHC during the Run 3 and 4~\cite{Citron:2018lsq}.

In addition, the coherent photoproduction of a resonance-like object with a mass around 1.7~GeV/$c^2$ which decays into a ${\pi^+\pi^-}$ pair is reported and compared with similar observations from other experiments.

\newenvironment{acknowledgement}{\relax}{\relax}
\begin{acknowledgement}
\section*{Acknowledgements}

The ALICE Collaboration would like to thank all its engineers and technicians for their invaluable contributions to the construction of the experiment and the CERN accelerator teams for the outstanding performance of the LHC complex.
The ALICE Collaboration gratefully acknowledges the resources and support provided by all Grid centres and the Worldwide LHC Computing Grid (WLCG) collaboration.
The ALICE Collaboration acknowledges the following funding agencies for their support in building and running the ALICE detector:
A. I. Alikhanyan National Science Laboratory (Yerevan Physics Institute) Foundation (ANSL), State Committee of Science and World Federation of Scientists (WFS), Armenia;
Austrian Academy of Sciences, Austrian Science Fund (FWF): [M 2467-N36] and Nationalstiftung f\"{u}r Forschung, Technologie und Entwicklung, Austria;
Ministry of Communications and High Technologies, National Nuclear Research Center, Azerbaijan;
Conselho Nacional de Desenvolvimento Cient\'{\i}fico e Tecnol\'{o}gico (CNPq), Financiadora de Estudos e Projetos (Finep), Funda\c{c}\~{a}o de Amparo \`{a} Pesquisa do Estado de S\~{a}o Paulo (FAPESP) and Universidade Federal do Rio Grande do Sul (UFRGS), Brazil;
Ministry of Education of China (MOEC) , Ministry of Science \& Technology of China (MSTC) and National Natural Science Foundation of China (NSFC), China;
Ministry of Science and Education and Croatian Science Foundation, Croatia;
Centro de Aplicaciones Tecnol\'{o}gicas y Desarrollo Nuclear (CEADEN), Cubaenerg\'{\i}a, Cuba;
Ministry of Education, Youth and Sports of the Czech Republic, Czech Republic;
Czech Science Foundation;
The Danish Council for Independent Research | Natural Sciences, the VILLUM FONDEN and Danish National Research Foundation (DNRF), Denmark;
Helsinki Institute of Physics (HIP), Finland;
Commissariat \`{a} l'Energie Atomique (CEA), Institut National de Physique Nucl\'{e}aire et de Physique des Particules (IN2P3) and Centre National de la Recherche Scientifique (CNRS) and R\'{e}gion des  Pays de la Loire, France;
Bundesministerium f\"{u}r Bildung und Forschung (BMBF) and GSI Helmholtzzentrum f\"{u}r Schwerionenforschung GmbH, Germany;
General Secretariat for Research and Technology, Ministry of Education, Research and Religions, Greece;
National Research, Development and Innovation Office, Hungary;
Department of Atomic Energy Government of India (DAE), Department of Science and Technology, Government of India (DST), University Grants Commission, Government of India (UGC) and Council of Scientific and Industrial Research (CSIR), India;
Indonesian Institute of Science, Indonesia;
Centro Fermi - Museo Storico della Fisica e Centro Studi e Ricerche Enrico Fermi and Istituto Nazionale di Fisica Nucleare (INFN), Italy;
Institute for Innovative Science and Technology , Nagasaki Institute of Applied Science (IIST), Japanese Ministry of Education, Culture, Sports, Science and Technology (MEXT) and Japan Society for the Promotion of Science (JSPS) KAKENHI, Japan;
Consejo Nacional de Ciencia (CONACYT) y Tecnolog\'{i}a, through Fondo de Cooperaci\'{o}n Internacional en Ciencia y Tecnolog\'{i}a (FONCICYT) and Direcci\'{o}n General de Asuntos del Personal Academico (DGAPA), Mexico;
Nederlandse Organisatie voor Wetenschappelijk Onderzoek (NWO), Netherlands;
The Research Council of Norway, Norway;
Commission on Science and Technology for Sustainable Development in the South (COMSATS), Pakistan;
Pontificia Universidad Cat\'{o}lica del Per\'{u}, Peru;
Ministry of Science and Higher Education and National Science Centre, Poland;
Korea Institute of Science and Technology Information and National Research Foundation of Korea (NRF), Republic of Korea;
Ministry of Education and Scientific Research, Institute of Atomic Physics and Ministry of Research and Innovation and Institute of Atomic Physics, Romania;
Joint Institute for Nuclear Research (JINR), Ministry of Education and Science of the Russian Federation, National Research Centre Kurchatov Institute, Russian Science Foundation and Russian Foundation for Basic Research, Russia;
Ministry of Education, Science, Research and Sport of the Slovak Republic, Slovakia;
National Research Foundation of South Africa, South Africa;
Swedish Research Council (VR) and Knut \& Alice Wallenberg Foundation (KAW), Sweden;
European Organization for Nuclear Research, Switzerland;
Suranaree University of Technology (SUT), National Science and Technology Development Agency (NSDTA) and Office of the Higher Education Commission under NRU project of Thailand, Thailand;
Turkish Atomic Energy Agency (TAEK), Turkey;
National Academy of  Sciences of Ukraine, Ukraine;
Science and Technology Facilities Council (STFC), United Kingdom;
National Science Foundation of the United States of America (NSF) and United States Department of Energy, Office of Nuclear Physics (DOE NP), United States of America.    
\end{acknowledgement}

\bibliographystyle{utphys}
\bibliography{RhoPbPb2015}

\newpage
\appendix
\section{The ALICE Collaboration}
\label{app:collab}

\begingroup
\small
\begin{flushleft}
S.~Acharya\Irefn{org142}\And 
D.~Adamov\'{a}\Irefn{org95}\And 
A.~Adler\Irefn{org74}\And 
J.~Adolfsson\Irefn{org81}\And 
M.M.~Aggarwal\Irefn{org100}\And 
G.~Aglieri Rinella\Irefn{org34}\And 
M.~Agnello\Irefn{org30}\And 
N.~Agrawal\Irefn{org10}\textsuperscript{,}\Irefn{org54}\And 
Z.~Ahammed\Irefn{org142}\And 
S.~Ahmad\Irefn{org16}
% ...
\And
\renewcommand\labelenumi{\textsuperscript{\theenumi}~}

\section*{Affiliation notes}
\renewcommand\theenumi{\roman{enumi}}
\begin{Authlist}
\item \Adef{org*}Deceased
\item \Adef{orgI}Italian National Agency for New Technologies, Energy and Sustainable Economic Development (ENEA), Bologna, Italy
\item \Adef{orgII}Dipartimento DET del Politecnico di Torino, Turin, Italy
\item \Adef{orgIII}M.V. Lomonosov Moscow State University, D.V. Skobeltsyn Institute of Nuclear, Physics, Moscow, Russia
\item \Adef{orgIV}Department of Applied Physics, Aligarh Muslim University, Aligarh, India
\item \Adef{orgV}Institute of Theoretical Physics, University of Wroclaw, Poland
\end{Authlist}

\section*{Collaboration Institutes}
\renewcommand\theenumi{\arabic{enumi}~}
\begin{Authlist}
\item \Idef{org1}A.I. Alikhanyan National Science Laboratory (Yerevan Physics Institute) Foundation, Yerevan, Armenia
\item \Idef{org2}Bogolyubov Institute for Theoretical Physics, National Academy of Sciences of Ukraine, Kiev, Ukraine
\item \Idef{org3}Bose Institute, Department of Physics  and Centre for Astroparticle Physics and Space Science (CAPSS), Kolkata, India
\item \Idef{org4}Budker Institute for Nuclear Physics, Novosibirsk, Russia
\item \Idef{org5}California Polytechnic State University, San Luis Obispo, California, United States
\item \Idef{org6}Central China Normal University, Wuhan, China
\item \Idef{org7}Centre de Calcul de l'IN2P3, Villeurbanne, Lyon, France
\item \Idef{org8}Centro de Aplicaciones Tecnol\'{o}gicas y Desarrollo Nuclear (CEADEN), Havana, Cuba
\item \Idef{org9}Centro de Investigaci\'{o}n y de Estudios Avanzados (CINVESTAV), Mexico City and M\'{e}rida, Mexico
\item \Idef{org10}Centro Fermi - Museo Storico della Fisica e Centro Studi e Ricerche ``Enrico Fermi', Rome, Italy
\item \Idef{org11}Chicago State University, Chicago, Illinois, United States
\item \Idef{org12}China Institute of Atomic Energy, Beijing, China
\item \Idef{org13}Comenius University Bratislava, Faculty of Mathematics, Physics and Informatics, Bratislava, Slovakia
\item \Idef{org14}COMSATS University Islamabad, Islamabad, Pakistan
\item \Idef{org15}Creighton University, Omaha, Nebraska, United States
\item \Idef{org16}Department of Physics, Aligarh Muslim University, Aligarh, India
\item \Idef{org17}Department of Physics, Pusan National University, Pusan, Republic of Korea
\item \Idef{org18}Department of Physics, Sejong University, Seoul, Republic of Korea
\item \Idef{org19}Department of Physics, University of California, Berkeley, California, United States
\item \Idef{org20}Department of Physics, University of Oslo, Oslo, Norway
\item \Idef{org21}Department of Physics and Technology, University of Bergen, Bergen, Norway
\item \Idef{org22}Dipartimento di Fisica dell'Universit\`{a} 'La Sapienza' and Sezione INFN, Rome, Italy
\item \Idef{org23}Dipartimento di Fisica dell'Universit\`{a} and Sezione INFN, Cagliari, Italy
\item \Idef{org24}Dipartimento di Fisica dell'Universit\`{a} and Sezione INFN, Trieste, Italy
\item \Idef{org25}Dipartimento di Fisica dell'Universit\`{a} and Sezione INFN, Turin, Italy
\item \Idef{org26}Dipartimento di Fisica e Astronomia dell'Universit\`{a} and Sezione INFN, Bologna, Italy
\item \Idef{org27}Dipartimento di Fisica e Astronomia dell'Universit\`{a} and Sezione INFN, Catania, Italy
\item \Idef{org28}Dipartimento di Fisica e Astronomia dell'Universit\`{a} and Sezione INFN, Padova, Italy
\item \Idef{org29}Dipartimento di Fisica `E.R.~Caianiello' dell'Universit\`{a} and Gruppo Collegato INFN, Salerno, Italy
\item \Idef{org30}Dipartimento DISAT del Politecnico and Sezione INFN, Turin, Italy
\item \Idef{org31}Dipartimento di Scienze e Innovazione Tecnologica dell'Universit\`{a} del Piemonte Orientale and INFN Sezione di Torino, Alessandria, Italy
\item \Idef{org32}Dipartimento di Scienze MIFT, Universit\`{a} di Messina, Messina, Italy
\item \Idef{org33}Dipartimento Interateneo di Fisica `M.~Merlin' and Sezione INFN, Bari, Italy
\item \Idef{org34}European Organization for Nuclear Research (CERN), Geneva, Switzerland
\item \Idef{org35}Faculty of Electrical Engineering, Mechanical Engineering and Naval Architecture, University of Split, Split, Croatia
\item \Idef{org36}Faculty of Engineering and Science, Western Norway University of Applied Sciences, Bergen, Norway
\item \Idef{org37}Faculty of Nuclear Sciences and Physical Engineering, Czech Technical University in Prague, Prague, Czech Republic
\item \Idef{org38}Faculty of Science, P.J.~\v{S}af\'{a}rik University, Ko\v{s}ice, Slovakia
\item \Idef{org39}Frankfurt Institute for Advanced Studies, Johann Wolfgang Goethe-Universit\"{a}t Frankfurt, Frankfurt, Germany
\item \Idef{org40}Fudan University, Shanghai, China
\item \Idef{org41}Gangneung-Wonju National University, Gangneung, Republic of Korea
\item \Idef{org42}Gauhati University, Department of Physics, Guwahati, India
\item \Idef{org43}Helmholtz-Institut f\"{u}r Strahlen- und Kernphysik, Rheinische Friedrich-Wilhelms-Universit\"{a}t Bonn, Bonn, Germany
\item \Idef{org44}Helsinki Institute of Physics (HIP), Helsinki, Finland
\item \Idef{org45}High Energy Physics Group,  Universidad Aut\'{o}noma de Puebla, Puebla, Mexico
\item \Idef{org46}Hiroshima University, Hiroshima, Japan
\item \Idef{org47}Hochschule Worms, Zentrum  f\"{u}r Technologietransfer und Telekommunikation (ZTT), Worms, Germany
\item \Idef{org48}Horia Hulubei National Institute of Physics and Nuclear Engineering, Bucharest, Romania
\item \Idef{org49}Indian Institute of Technology Bombay (IIT), Mumbai, India
\item \Idef{org50}Indian Institute of Technology Indore, Indore, India
\item \Idef{org51}Indonesian Institute of Sciences, Jakarta, Indonesia
\item \Idef{org52}INFN, Laboratori Nazionali di Frascati, Frascati, Italy
\item \Idef{org53}INFN, Sezione di Bari, Bari, Italy
\item \Idef{org54}INFN, Sezione di Bologna, Bologna, Italy
\item \Idef{org55}INFN, Sezione di Cagliari, Cagliari, Italy
\item \Idef{org56}INFN, Sezione di Catania, Catania, Italy
\item \Idef{org57}INFN, Sezione di Padova, Padova, Italy
\item \Idef{org58}INFN, Sezione di Roma, Rome, Italy
\item \Idef{org59}INFN, Sezione di Torino, Turin, Italy
\item \Idef{org60}INFN, Sezione di Trieste, Trieste, Italy
\item \Idef{org61}Inha University, Incheon, Republic of Korea
\item \Idef{org62}Institute for Nuclear Research, Academy of Sciences, Moscow, Russia
\item \Idef{org63}Institute for Subatomic Physics, Utrecht University/Nikhef, Utrecht, Netherlands
\item \Idef{org64}Institute of Experimental Physics, Slovak Academy of Sciences, Ko\v{s}ice, Slovakia
\item \Idef{org65}Institute of Physics, Homi Bhabha National Institute, Bhubaneswar, India
\item \Idef{org66}Institute of Physics of the Czech Academy of Sciences, Prague, Czech Republic
\item \Idef{org67}Institute of Space Science (ISS), Bucharest, Romania
\item \Idef{org68}Institut f\"{u}r Kernphysik, Johann Wolfgang Goethe-Universit\"{a}t Frankfurt, Frankfurt, Germany
\item \Idef{org69}Instituto de Ciencias Nucleares, Universidad Nacional Aut\'{o}noma de M\'{e}xico, Mexico City, Mexico
\item \Idef{org70}Instituto de F\'{i}sica, Universidade Federal do Rio Grande do Sul (UFRGS), Porto Alegre, Brazil
\item \Idef{org71}Instituto de F\'{\i}sica, Universidad Nacional Aut\'{o}noma de M\'{e}xico, Mexico City, Mexico
\item \Idef{org72}iThemba LABS, National Research Foundation, Somerset West, South Africa
\item \Idef{org73}Jeonbuk National University, Jeonju, Republic of Korea
\item \Idef{org74}Johann-Wolfgang-Goethe Universit\"{a}t Frankfurt Institut f\"{u}r Informatik, Fachbereich Informatik und Mathematik, Frankfurt, Germany
\item \Idef{org75}Joint Institute for Nuclear Research (JINR), Dubna, Russia
\item \Idef{org76}Korea Institute of Science and Technology Information, Daejeon, Republic of Korea
\item \Idef{org77}KTO Karatay University, Konya, Turkey
\item \Idef{org78}Laboratoire de Physique des 2 Infinis, Irène Joliot-Curie, Orsay, France
\item \Idef{org79}Laboratoire de Physique Subatomique et de Cosmologie, Universit\'{e} Grenoble-Alpes, CNRS-IN2P3, Grenoble, France
\item \Idef{org80}Lawrence Berkeley National Laboratory, Berkeley, California, United States
\item \Idef{org81}Lund University Department of Physics, Division of Particle Physics, Lund, Sweden
\item \Idef{org82}Nagasaki Institute of Applied Science, Nagasaki, Japan
\item \Idef{org83}Nara Women{'}s University (NWU), Nara, Japan
\item \Idef{org84}National and Kapodistrian University of Athens, School of Science, Department of Physics , Athens, Greece
\item \Idef{org85}National Centre for Nuclear Research, Warsaw, Poland
\item \Idef{org86}National Institute of Science Education and Research, Homi Bhabha National Institute, Jatni, India
\item \Idef{org87}National Nuclear Research Center, Baku, Azerbaijan
\item \Idef{org88}National Research Centre Kurchatov Institute, Moscow, Russia
\item \Idef{org89}Niels Bohr Institute, University of Copenhagen, Copenhagen, Denmark
\item \Idef{org90}Nikhef, National institute for subatomic physics, Amsterdam, Netherlands
\item \Idef{org91}NRC Kurchatov Institute IHEP, Protvino, Russia
\item \Idef{org92}NRC «Kurchatov Institute»  - ITEP, Moscow, Russia
\item \Idef{org93}NRNU Moscow Engineering Physics Institute, Moscow, Russia
\item \Idef{org94}Nuclear Physics Group, STFC Daresbury Laboratory, Daresbury, United Kingdom
\item \Idef{org95}Nuclear Physics Institute of the Czech Academy of Sciences, \v{R}e\v{z} u Prahy, Czech Republic
\item \Idef{org96}Oak Ridge National Laboratory, Oak Ridge, Tennessee, United States
\item \Idef{org97}Ohio State University, Columbus, Ohio, United States
\item \Idef{org98}Petersburg Nuclear Physics Institute, Gatchina, Russia
\item \Idef{org99}Physics department, Faculty of science, University of Zagreb, Zagreb, Croatia
\item \Idef{org100}Physics Department, Panjab University, Chandigarh, India
\item \Idef{org101}Physics Department, University of Jammu, Jammu, India
\item \Idef{org102}Physics Department, University of Rajasthan, Jaipur, India
\item \Idef{org103}Physikalisches Institut, Eberhard-Karls-Universit\"{a}t T\"{u}bingen, T\"{u}bingen, Germany
\item \Idef{org104}Physikalisches Institut, Ruprecht-Karls-Universit\"{a}t Heidelberg, Heidelberg, Germany
\item \Idef{org105}Physik Department, Technische Universit\"{a}t M\"{u}nchen, Munich, Germany
\item \Idef{org106}Politecnico di Bari, Bari, Italy
\item \Idef{org107}Research Division and ExtreMe Matter Institute EMMI, GSI Helmholtzzentrum f\"ur Schwerionenforschung GmbH, Darmstadt, Germany
\item \Idef{org108}Rudjer Bo\v{s}kovi\'{c} Institute, Zagreb, Croatia
\item \Idef{org109}Russian Federal Nuclear Center (VNIIEF), Sarov, Russia
\item \Idef{org110}Saha Institute of Nuclear Physics, Homi Bhabha National Institute, Kolkata, India
\item \Idef{org111}School of Physics and Astronomy, University of Birmingham, Birmingham, United Kingdom
\item \Idef{org112}Secci\'{o}n F\'{\i}sica, Departamento de Ciencias, Pontificia Universidad Cat\'{o}lica del Per\'{u}, Lima, Peru
\item \Idef{org113}St. Petersburg State University, St. Petersburg, Russia
\item \Idef{org114}Stefan Meyer Institut f\"{u}r Subatomare Physik (SMI), Vienna, Austria
\item \Idef{org115}SUBATECH, IMT Atlantique, Universit\'{e} de Nantes, CNRS-IN2P3, Nantes, France
\item \Idef{org116}Suranaree University of Technology, Nakhon Ratchasima, Thailand
\item \Idef{org117}Technical University of Ko\v{s}ice, Ko\v{s}ice, Slovakia
\item \Idef{org118}Technische Universit\"{a}t M\"{u}nchen, Excellence Cluster 'Universe', Munich, Germany
\item \Idef{org119}The Henryk Niewodniczanski Institute of Nuclear Physics, Polish Academy of Sciences, Cracow, Poland
\item \Idef{org120}The University of Texas at Austin, Austin, Texas, United States
\item \Idef{org121}Universidad Aut\'{o}noma de Sinaloa, Culiac\'{a}n, Mexico
\item \Idef{org122}Universidade de S\~{a}o Paulo (USP), S\~{a}o Paulo, Brazil
\item \Idef{org123}Universidade Estadual de Campinas (UNICAMP), Campinas, Brazil
\item \Idef{org124}Universidade Federal do ABC, Santo Andre, Brazil
\item \Idef{org125}University of Cape Town, Cape Town, South Africa
\item \Idef{org126}University of Houston, Houston, Texas, United States
\item \Idef{org127}University of Jyv\"{a}skyl\"{a}, Jyv\"{a}skyl\"{a}, Finland
\item \Idef{org128}University of Liverpool, Liverpool, United Kingdom
\item \Idef{org129}University of Science and Technology of China, Hefei, China
\item \Idef{org130}University of South-Eastern Norway, Tonsberg, Norway
\item \Idef{org131}University of Tennessee, Knoxville, Tennessee, United States
\item \Idef{org132}University of the Witwatersrand, Johannesburg, South Africa
\item \Idef{org133}University of Tokyo, Tokyo, Japan
\item \Idef{org134}University of Tsukuba, Tsukuba, Japan
\item \Idef{org135}Universit\'{e} Clermont Auvergne, CNRS/IN2P3, LPC, Clermont-Ferrand, France
\item \Idef{org136}Universit\'{e} de Lyon, Universit\'{e} Lyon 1, CNRS/IN2P3, IPN-Lyon, Villeurbanne, Lyon, France
\item \Idef{org137}Universit\'{e} de Strasbourg, CNRS, IPHC UMR 7178, F-67000 Strasbourg, France, Strasbourg, France
\item \Idef{org138}Universit\'{e} Paris-Saclay Centre d'Etudes de Saclay (CEA), IRFU, D\'{e}partment de Physique Nucl\'{e}aire (DPhN), Saclay, France
\item \Idef{org139}Universit\`{a} degli Studi di Foggia, Foggia, Italy
\item \Idef{org140}Universit\`{a} degli Studi di Pavia, Pavia, Italy
\item \Idef{org141}Universit\`{a} di Brescia, Brescia, Italy
\item \Idef{org142}Variable Energy Cyclotron Centre, Homi Bhabha National Institute, Kolkata, India
\item \Idef{org143}Warsaw University of Technology, Warsaw, Poland
\item \Idef{org144}Wayne State University, Detroit, Michigan, United States
\item \Idef{org145}Westf\"{a}lische Wilhelms-Universit\"{a}t M\"{u}nster, Institut f\"{u}r Kernphysik, M\"{u}nster, Germany
\item \Idef{org146}Wigner Research Centre for Physics, Budapest, Hungary
\item \Idef{org147}Yale University, New Haven, Connecticut, United States
\item \Idef{org148}Yonsei University, Seoul, Republic of Korea
\end{Authlist}
\endgroup
\end{document}